\documentclass[runningheads]{llncs}

\usepackage{etex}
\usepackage{tikz}
\usetikzlibrary{calc} 						
\usetikzlibrary{shapes,arrows}				
\usetikzlibrary{shapes.multipart}			
\usetikzlibrary{positioning}
\usepackage[all,cmtip]{xy}					
\usepackage{tikz-qtree,tikz-qtree-compat}	

\usepackage{url}
\usepackage{turnstile}
\usepackage{epsfig}
\usepackage{amssymb}
\usepackage[all]{xypic}
\usepackage{amsmath}
\usepackage{latexsym}
\usepackage{cancel}

\long\def\comment#1{}

\newcommand{\SAlgebra}{\cA}
\newcommand{\SStructure}{\cA}

\newcommand{\SModel}{\cA}
\newcommand{\InitialModel}{\cI}

\newcommand{\SemDomain}{\cA}
\newcommand{\StructDomain}{\cA}
\newcommand{\ModelDomain}{\cA}

\newcommand{\SurHomTh}{\mt{SuH}}

\newcommand{\IF}{\Leftarrow}

\newcommand{\AGES}{{\sf AGES}}
\newcommand{\AltErgo}{{\sf Alt-Ergo}}

\newcommand{\Maude}{{\sf Maude}}
\newcommand{\MaceFour}{{\sf Mace4}}

\newcommand{\PDLtableau}{{\sf PDL-tableau}}
\newcommand{\ProverNine}{{\sf Prover9}}
\newcommand{\Princess}{{\sf Princess}}
\newcommand{\SEM}{{\sf SEM}}

\newcommand{\bigfracn}[3]{
\begin{array}[b]{c}
\displaystyle #1 \\\hline\displaystyle #2 
\end{array}
\hbox to 0pt{\raisebox{0.7em}{{\tiny (#3)}}}
}

\makeatletter
\newenvironment{prog}{\vspace{0.7ex}\par
\setlength{\parindent}{0.7cm}
\obeylines\@vobeyspaces\tt}{\vspace{0.0ex}\noindent
}
\makeatother
\newcommand{\startprog}{\begin{prog}}
\newcommand{\stopprog}{\end{prog}\noindent}

\makeatletter
\newenvironment{smallprog}{\vspace{0.7ex}\par
\setlength{\parindent}{0.7cm}
\obeylines\@vobeyspaces\tt\small}{\vspace{0.7ex}\noindent
}
\makeatother
\newcommand{\fstartprog}{\begin{smallprog}}
\newcommand{\fstopprog}{\end{smallprog}\noindent}

\makeatletter
\newenvironment{nismallprog}{\vspace{0.7ex}\par
\setlength{\parindent}{0.0cm}
\obeylines\@vobeyspaces\tt\small}{\vspace{0.7ex}\noindent
}
\makeatother
\newcommand{\fnistartprog}{\begin{nismallprog}}
\newcommand{\fnistopprog}{\end{nismallprog}\noindent}

\newcommand{\pr}[1]{\mathtt{\tt #1}}   
 
\newcommand{\mt}[1]{\mbox{\sf #1}}   

\newcommand{\ol}[1]{\overline{#1}}

\def\defemb#1#2{\expandafter\def\csname #1\endcsname
{\relax\ifmmode #2\else\hbox{$#2$}\fi}}
\defemb{cA}{{\cal A}}
\defemb{cB}{{\cal B}}
\defemb{cC}{{\cal C}}
\defemb{cD}{{\cal D}}
\defemb{cE}{{\cal E}}
\defemb{cF}{{\cal F}}
\defemb{cG}{{\cal G}}
\defemb{cH}{{\cal H}}
\defemb{cI}{{\cal I}}
\defemb{cJ}{{\cal J}}
\defemb{cL}{{\cal L}}
\defemb{cM}{{\cal M}}
\defemb{cN}{{\cal N}}
\defemb{cO}{{\cal O}}
\defemb{cP}{{\cal P}}
\defemb{cQ}{{\cal Q}}
\defemb{cR}{{\cal R}}
\defemb{cS}{{\cal S}}
\defemb{cT}{{\cal T}}
\defemb{cU}{{\cal U}}
\defemb{cW}{{\cal W}}
\defemb{cV}{{\cal V}}
\defemb{cX}{{\cal X}}
\defemb{cZ}{{\cal Z}}

\newcommand{\formulasOn}[1]{\mathit{Form}_{#1}}
\newcommand{\SPformulas}{\formulasOn{\SPSignature}}

\newcommand{\SSignature}{{\Sigma}}
\newcommand{\SPSignature}{{\Omega}}

\newcommand{\SSymbols}{{\Sigma}}
\newcommand{\SPredicates}{{\Pi}}
\newcommand{\STermsOn}[2]{{\cT_{#1}(#2)}}
\newcommand{\STerms}{{\STermsOn{\SSymbols}{\Variables}}}
\newcommand{\GSTermsOn}[1]{{\cT_{#1}}}
\newcommand{\GSTerms}{{\GSTermsOn{\SSymbols}}}

\newcommand{\CSymbols}{{\cC}}
\newcommand{\DSymbols}{{\cD}}

\newcommand{\Variables}{{\cX}}

\newcommand{\toppos}{\Lambda}

\newcommand{\nat}{\mathbb{N}}
\newcommand{\naturals}{\mathbb{N}}

\newcommand{\integers}{\mathbb{Z}}

\newcommand{\exr}{\stackrel{\toppos}{\to}}

\newcommand{\genSort}{\mathtt{S}}

\newcommand{\SortNat}{\mathit{Nat}}

\newcommand{\exSymbType}[1]{\mathsf{#1}}

\newcommand{\Fa}{\exSymbType{a}}
\newcommand{\Fadd}{\exSymbType{add}}

\newcommand{\Fb}{\exSymbType{b}}
\newcommand{\Fc}{\exSymbType{c}}

\newcommand{\Fd}{\exSymbType{d}}

\newcommand{\Ff}{\exSymbType{f}}

\newcommand{\Fg}{\exSymbType{g}}

\newcommand{\Fmul}{\exSymbType{mul}}

\newcommand{\Fnil}{\exSymbType{nil}}
\newcommand{\Fnon}{\exSymbType{non}}

\newcommand{\Fs}{\exSymbType{s}}

\newcommand{\Fsk}{\exSymbType{sk}}

\newcommand{\Fz}{\exSymbType{0}}

\tikzstyle{decision} = [diamond, draw, fill=yellow!20, text width=5em, text badly centered, minimum height=4em, inner sep=0pt, aspect=2]
\tikzstyle{block} = [rectangle, draw,fill=blue!20, text width=5em, text centered, minimum height=4em, rounded corners]
\tikzstyle{cloud} = [ellipse, draw,fill=red!20, text width=5em, text centered, minimum height=4em]
\tikzstyle{line} = [draw, -latex']
\tikzstyle{blockR} = [rectangle, draw, fill=red!20, text centered, minimum height=4em, rounded corners, minimum height=0.75cm]
\tikzstyle{blockB} = [rectangle, draw, fill=blue!20, text centered, minimum height=4em, rounded corners, minimum height=0.75cm]
\tikzstyle{blockG} = [rectangle, draw, fill=green!20, text centered, minimum height=4em, rounded corners, minimum height=0.75cm]
\tikzstyle{blockY} = [rectangle, draw, fill=yellow!20, text centered, minimum height=4em, rounded corners, minimum height=0.75cm]
\tikzstyle{blockW} = [rectangle, draw, fill=white!20, text centered, minimum height=4em, rounded corners, minimum height=0.75cm]
\tikzstyle{blockK} = [rectangle, draw, fill=gray!20, text centered, minimum height=4em, rounded corners, minimum height=0.75cm]
\tikzstyle{mblockB} = [rectangle, draw, fill=blue!20, text centered, minimum height=4em, double,rounded corners, minimum height=0.75cm]
\tikzstyle{mblockW} = [rectangle, draw, fill=white!20, text centered, minimum height=4em, double,rounded corners, minimum height=0.75cm]
\tikzstyle{mblockR} = [rectangle, draw, fill=red!20, text centered, minimum height=4em, double,rounded corners, minimum height=0.75cm]
\tikzstyle{mblockG} = [rectangle, draw, fill=green!20, text centered, minimum height=4em, double,rounded corners, minimum height=0.75cm]

\tikzstyle{circleW} = [circle, draw, fill=white!20, text centered, minimum height=4em, rounded corners, minimum height=0.75cm]
\tikzstyle{triangleW} = [isosceles triangle, draw, fill=white!20, text centered, shape border rotate = -90, isosceles triangle stretches]
\tikzstyle{triangleG} = [isosceles triangle, draw, fill=green!20, text centered, shape border rotate = -90, isosceles triangle stretches]
\tikzstyle{triangleR} = [isosceles triangle, draw, fill=red!20, text centered, shape border rotate = -90, isosceles triangle stretches]

\date{}

\begin{document}

\title{Proving Program Properties as First-Order Satisfiability%
\thanks{Partially supported by the EU (FEDER), 
projects 
TIN2015-69175-C4-1-R,
and GV PROMETEOII/2015/013.}}

\author{
  Salvador Lucas
}

\institute{
  DSIC, Universitat Polit\`ecnica de Val\`encia, Spain,\\
  \url{http://slucas.webs.upv.es/}
           }
\titlerunning{Proving Program Properties as First-Order Satisfiability}

\authorrunning{Lucas}

\maketitle

\begin{abstract}
Program semantics can often be expressed as a (many-sorted) first-order theory $\cS$, and program properties
as sentences $\varphi$ which are intended to hold in the \emph{canonical model} of such a theory, which is often incomputable.
Recently, we have shown that properties $\varphi$ expressed as the existential closure of a boolean combination of atoms
can be \emph{disproved} by just finding a model of $\cS$ and the \emph{negation} $\neg\varphi$ of $\varphi$. 
Furthermore, this idea works quite well in practice due to the existence of powerful tools for the automatic generation of
models for (many-sorted) first-order theories.
In this paper we extend our previous results to \emph{arbitrary} properties, expressed as 
sentences without any special restriction.
Consequently, one can prove a program property $\varphi$ by just \emph{finding a model} of an appropriate 
theory (including $\cS$ and possibly something else) and an appropriate first-order formula related to $\varphi$.
Beyond its possible theoretical interest, we show that our results can also be of practical use in several respects.
\end{abstract}


\setcounter{page}{1}

\section{Introduction}
\label{SecIntroduction}

Given a first-order theory $\cS$ and a sentence $\varphi$, finding a model $\SStructure$ of $\cS\:\cup\:\{\neg\varphi\}$, i.e., such that 
$\SStructure\models\cS\cup\{\neg\varphi\}$ holds, shows indeed that 
$\varphi$ is \emph{not} a logical consequence of $\cS$: 
there is at least one model of $\cS$ (e.g., $\SStructure$) which does not
satisfy $\varphi$ (as it satisfies $\neg\varphi$).
Provability of $\varphi$ in $\cS$, i.e., $\cS\vdash\varphi$, implies (by correctness of the proof calculus) that $\varphi$ is a logical consequence of $\cS$
 (written $\cS\models\varphi$).
Thus, $\SStructure\models\cS\cup\{\neg\varphi\}$ \emph{disproves} $\varphi$ regarding $\cS$;
this can be written $\neg(\cS\vdash\varphi)$ 
by using some metalevel notation.
In general, this does not allow us to conclude that $\neg\varphi$ is proved, i.e., $\cS\vdash\neg\varphi$, or is a logical consequence of $\cS$, i.e., $\cS\models\neg\varphi$.
What can be concluded about $\neg\varphi$ regarding $\cS$ 
from the fact that  $\SStructure\models\cS\cup\{\neg\varphi\}$ holds?
Can this be advantageously used in a `logic-based' approach to program analysis?

In \cite{Lucas_AnalysisOfRewritingBasedSystemsAsFirstOrderTheories_LNCS_LOPSTR17}, some answers 
to these questions are given:
a sentence $\varphi$ which is an \emph{Existentially Closed Boolean 
Combination of Atoms} (ECBCA for short)
does \emph{not} hold in the initial model $\cI_\cS$ of a theory $\cS$ consisting of a set of ground atoms
if we find a model $\SStructure$ of $\cS\cup\{\neg\varphi\}$ 
\cite[Corollary 2]{Lucas_AnalysisOfRewritingBasedSystemsAsFirstOrderTheories_LNCS_LOPSTR17}.
This is useful in program analysis when considering 
programs $P$ that are given a theory $\ol{P}$ representing 
its operational semantics so that the execution of $P$ is described as a set $\cI_P$ of (ground) atoms
 $A$ which can be proved from $\ol{P}$ (i.e., $\cI_P$ is the \emph{initial} model of
 $\ol{P}$ in the usual first-order sense; in the following, we often refer to it as its \emph{canonical} model
 \cite[Section 1.5]{Hodges_ModelTheory_1993}). Actually, rather than being logical consequences of $\ol{P}$, 
 the intended meaning of first-order sentences $\varphi$ that represent properties of $P$ is that they 
hold in the \emph{initial model} of $\ol{P}$, 
see \cite[Chapter 4]{Clark_PredicateLogicAsAComputationalFormalism_TR79}, for instance.

In \cite{Lucas_AnalysisOfRewritingBasedSystemsAsFirstOrderTheories_LNCS_LOPSTR17,LucGut_UseOfLogicalModelsForProvingInfeasibilityInTermRewriting_IPL18} 
we applied this approach to prove computational properties of rewriting-based
systems in practice.
This includes 
 Term Rewriting Systems (TRSs \cite{BaaNip_TermRewAllThat_1998}) 
and more general rewriting-based formalisms 
\cite{BruMes_SemFoundGRT_TCS06,%
GogMes_ModelsAndEqualityForLogicalProgramming_TAPSOFT87,%
Meseguer_20YearsRewLogic_JLAP12,Ohlebusch_AdvTopicsTermRew_2002}.

\begin{example}\label{ExAddMul}\label{ExAddMul_DoubleIsNotOdd}
Consider the following TRS $\cR$ with the well-known rules defining the addition and product of natural numbers in Peano's notation:
\\[-0.5cm]
\begin{tabular}{cc}
\hspace{-1cm}
\begin{minipage}[t]{.55\linewidth}
\begin{eqnarray}
\Fadd(\Fz,x) & \to &x\label{ExAddMul_rule1}\\
\Fadd(\Fs(x),y) & \to & \Fs(\Fadd(x,y))\label{ExAddMul_rule2}
\end{eqnarray}\nonumber
\end{minipage} & 
\begin{minipage}[t]{.51\linewidth}
\begin{eqnarray}
\Fmul(\Fz,x) & \to &\Fz\label{ExAddMul_rule3}\\
\Fmul(\Fs(x),y) & \to & \Fadd(y,\Fmul(x,y))\label{ExAddMul_rule4}
\end{eqnarray}\nonumber
\end{minipage}
\end{tabular}\\[0.3cm]
The associated theory $\ol{\cR}$ is the following:\\[-0.8cm] 

{\small 
\begin{tabular}{c@{\hspace{-0.7cm}}c}
\hspace{-0.95cm}
\begin{minipage}[t]{.57\linewidth}
\begin{eqnarray}
(\forall x)\: x & \to^* & x\nonumber\label{ExAddMul_HornTh_reflexivity}\\
(\forall x,y,z)\: x \to y \wedge y \to^* z \Rightarrow x & \to^* & z\nonumber\label{ExAddMul_HornTh_transitivity}\\
(\forall x,y)\: x \to y  \Rightarrow \Fs(x) & \to & \Fs(y)\nonumber\label{ExAddMul_HornTh_congruenceSucc}\\
(\forall x,y,z)\: x \to y  \Rightarrow \Fadd(x,z) & \to & \Fadd(y,z)\nonumber\label{ExAddMul_HornTh_congruenceAdd1}\\
(\forall x,y,z)\: x \to y  \Rightarrow \Fadd(z,x) & \to & \Fadd(z,y)\nonumber\label{ExAddMul_HornTh_congruenceAdd2}\\
(\forall x,y,z)\: x \to y  \Rightarrow \Fmul(x,z) & \to & \Fmul(y,z)\nonumber\label{ExAddMul_HornTh_congruenceMul1}\\
(\forall x,y,z)\: x \to y  \Rightarrow \Fmul(z,x) & \to & \Fmul(z,y)\nonumber\label{ExAddMul_HornTh_congruenceMul2}
\end{eqnarray}\nonumber
\end{minipage} & 
\begin{minipage}[t]{.56\linewidth}
\begin{eqnarray}
(\forall x)\: \Fadd(\Fz,x) & \to &x \nonumber\label{ExAddMul_HornTh_rule1}\\
(\forall x,y)\: \Fadd(\Fs(x),y) & \to & \Fs(\Fadd(x,y)) \nonumber\label{ExLoopingnessInCTRSs_HornTh_rule2}\\
(\forall x)\: \Fmul(\Fz,x) & \to &\Fz\nonumber \label{ExAddMul_HornTh_rule3}\\
(\forall x,y)\:\Fmul(\Fs(x),y) & \to & \Fadd(y,\Fmul(x,y)) \nonumber\label{ExLoopingnessInCTRSs_HornTh_rule4}
\end{eqnarray}\nonumber
\end{minipage}
\end{tabular}}\\[0.2cm]

\noindent
The first sentence in the first column 
represents \emph{reflexivity} of many-step rewriting, with predicate symbol $\to^*$;
the second sentence 
shows how one-step rewriting, with predicate symbol $\to$, contributes to $\to^*$.
The next sentences describe the \emph{propagation} of rewriting steps to (arguments of) symbols $\Fs$, $\Fadd$ and $\Fmul$.
The second column describes the rules of $\cR$.
More details can be found in  \cite[Section 4]{Lucas_AnalysisOfRewritingBasedSystemsAsFirstOrderTheories_LNCS_LOPSTR17}.
In the initial or \emph{least Herbrand model} $\cI_\cR$ 
of $\ol{\cR}$, $\to$ and $\to^*$ are interpreted as the 
sets $(\to)^{\cI_\cR}$ and $(\to^*)^{\cI_\cR}$ of all pairs $(s,t)$ of ground terms $s$ and $t$ such that $s\to_\cR t$ and $s\to^*_\cR t$, respectively.
Now, we can express the property ``\emph{the double of some natural number can be
an odd number}'' as an ECBCA:
\begin{eqnarray}
(\exists x) (\exists y) (\exists z)~\Fadd(x,x) \to^*z\wedge \Fs(\Fmul(\Fs(\Fs(\Fz)),y))\to^*z\label{PropExAddMul_DoubleIsNotOdd}
\end{eqnarray}
With the automatic model generator \MaceFour\ \cite{McCune_Prove9andMace4_Unpublished10} we find a model of 
$\ol{\cR}\:\cup\:\{\neg(\ref{PropExAddMul_DoubleIsNotOdd})\}$ with domain $\SStructure=\{0,1\}$.
Function symbols are interpreted as follows:
$\Fz^\SStructure = 0$; 
$\Fs^\SStructure(x)=1-x$;
$\Fadd^\SStructure(x,y)$ returns $0$ if $x=y$ and $1$ otherwise;
$\Fmul^\SStructure(x,y)$ returns $1$ if $x=y=1$ and $0$ otherwise.
Predicates $\to$ and $\to^*$ are both interpreted as the \emph{equality}.
Thus, we have \emph{proved} that (\ref{PropExAddMul_DoubleIsNotOdd}) does \emph{not} hold for $\cR$.
\end{example}
Our approach in \cite{Lucas_AnalysisOfRewritingBasedSystemsAsFirstOrderTheories_LNCS_LOPSTR17} relies on
the notion of \emph{preservation} of 
a formula under homomorphisms
$h$ 
between interpretations. 
Roughly speaking, a homomorphism $h$ \emph{preserves} a formula $\varphi$
if  $\varphi$ is satisfied in the target interpretation of $h$ whenever $\varphi$ is satisfied in its domain interpretation 
\cite[Section 2.4]{Hodges_ModelTheory_1993}.
Homomorphisms preserve ECBCA \cite[Theorem 2.4.3(a)]{Hodges_ModelTheory_1993};
the results in \cite{Lucas_AnalysisOfRewritingBasedSystemsAsFirstOrderTheories_LNCS_LOPSTR17} rely on this fact.
In this paper we \emph{extend} \cite{Lucas_AnalysisOfRewritingBasedSystemsAsFirstOrderTheories_LNCS_LOPSTR17} 
to deal with more general program properties.
Homomorphisms preserve other first-order sentences if further requirements are imposed: 
(i) \emph{positive} sentences (where connective `$\neg$' is absent) are preserved under
\emph{surjective} homomorphisms and 
(ii) \emph{arbitrary} sentences are preserved under \emph{embeddings}
 \cite[Theorem 2.4.3]{Hodges_ModelTheory_1993}.
In contrast to \cite{Lucas_AnalysisOfRewritingBasedSystemsAsFirstOrderTheories_LNCS_LOPSTR17} (and  \cite{Hodges_ModelTheory_1993}), 
here we focus on \emph{many-sorted logic} \cite{Wang_LogicOfManySortedTheories_JSL52} (see Section  \ref{SecOrderSortedFirstOrderLogic}). 
This has an important advantage: since homomorphisms in many-sorted
logic with set of sorts $S$ are actually a \emph{family} $h_s$ of homomorphisms between components of sort $s$ 
for each $s\in S$, 
the preservation requirements for $h_s$ depend on the \emph{specific quantification} of variables $x:s$ \emph{for such a sort}.
In Section \ref{SecPreservationManySorted} we provide a \emph{unique} preservation theorem that subsumes the results in \cite{Lucas_AnalysisOfRewritingBasedSystemsAsFirstOrderTheories_LNCS_LOPSTR17},
and even improves \cite{Hodges_ModelTheory_1993}.
Section \ref{SecSurjectiveHomomorphisms} investigates how to guarantee surjectivity of homomorphisms.
Section \ref{SecExampleOfApplication} shows several application examples taken from
Table \ref{TableSomeFOpropertiesRewritingBasedSystems}, which shows some properties
of rewriting-based systems that could not be captured in \cite{Lucas_AnalysisOfRewritingBasedSystemsAsFirstOrderTheories_LNCS_LOPSTR17}
but we are able to handle now. Here,
$t(\vec{x})$ is a term with variables $\vec{x}$ (or just $t$ if it is \emph{ground}), 
$\CSymbols$ (and $\DSymbols$) are the \emph{constructor} (resp.\ \emph{defined}) symbols in the TRS,
and $\exr$ is \emph{topmost rewriting}.
\begin{table}[t]
\begin{center}
\begin{tabular}{|@{~}c@{~}|@{~}c@{~}|}
\hline
\bf \emph{Property} & $\varphi$ \\
\hline
\emph{G}round \emph{red}ucible & $(\forall\vec{x})\:(\exists y)~t(\vec{x})\to y$ \\
\emph{Comp}letely defined symbol $f$ & $(\forall\vec{x})(\exists y)~f(x_1,\ldots,x_k)\to y$\\
\emph{Comp}letely defined TRS & $(\forall\vec{x})(\exists\vec{y})\bigwedge_{f\in\DSymbols}~f(x_1,\ldots,x_{ar(f)})\to y_f$\\
\emph{Prod}uctive & $(\forall x)(\exists\vec{y})~\bigvee_{c\in\CSymbols}x\to^*c(y_1,\ldots,y_k)$\\
\emph{Nont}erminating & $(\exists x)(\forall n\in\naturals)(\exists y)~x\to^ny$\\
\emph{Inf}initely root-reducible & $(\exists x)(\forall n\in\naturals)(\exists y)~x(\to^*\circ\exr)^ny$\\
\hline
\emph{Norm}alizing term & $(\exists x)\:(t\to^* x\wedge \neg(\exists y)~x\to y)$\\
\emph{Norm}alizing TRS (WN) & $(\forall x)(\exists y)\:(x\to^* y\wedge \neg(\exists z)~y\to z)$\\
\emph{Locally} confluent (WCR) & $(\forall x,y,z)~x\to y\wedge x\to z \Rightarrow (\exists u)~x\to^*u\wedge z\to^*u$\\
\emph{Conf}luent (CR) &  $(\forall x,y,z)~x\to^* y\wedge x\to^* z \Rightarrow (\exists u)~x\to^*u\wedge z\to^*u$\\
\hline
\end{tabular}
\end{center}
\caption{Some properties about rewriting-based systems}
\label{TableSomeFOpropertiesRewritingBasedSystems}
\end{table}
Section \ref{SecWitnessesOfSemanticRefutation} discusses the possibility of providing more information about
\emph{disproved} properties by means of \emph{refutation witnesses}, i.e., (counter)examples of \emph{sentences} 
which are synthesized from the models that are used to disprove the property.
Section \ref{SecDealingWithGeneralSentences} shows how to deal with completely general sentences by means of a
simple example.
Section \ref{SecRelatedWork} discusses some related work. Section \ref{SecConclusions} concludes.

\section{Many-Sorted First-Order Logic}\label{SecOrderSortedFirstOrderLogic}

Given a set of \emph{sorts} $S$, a \emph{(many-sorted) signature (with predicates)} 
$\SPSignature=(S,\SSymbols,\SPredicates)$ consists of a set of sorts $S$,
an $S^{\ast} \times S$-indexed family of sets $\Sigma =
\{\Sigma_{w,s}\}_{(w,s) \in S^* \times S}$ containing
\emph{function symbols} 
$f \in \Sigma_{s_{1}\cdots s_{k},s}$, with a rank declaration $f: s_{1}\cdots s_{k} \to s$
(constant symbols $c$ 
have rank declaration $c:\lambda\to s$, where $\lambda$ 
denotes the \emph{empty} sequence),
and an $S^+$-indexed family of sets $\SPredicates=\{\SPredicates_w\}_{w\in S^+}$ of ranked predicates 
$P:w$.
Given an $S$-sorted set $\Variables=\{\Variables_s\mid s\in S\}$ of \emph{mutually disjoint}
sets of variables (which are also disjoint from $\SSymbols$), 
the set $\STerms_s$ of terms of sort $s$ is the least set such that
$\Variables_{s}\subseteq\STerms_s$
and 
for each $f: s_{1}\ldots s_{k} \rightarrow s$ 
and $t_i\in\STerms_{s_i}$, $1\leq i\leq k$,  $f(t_1,\ldots,t_k)\in\STerms_s$. 
If $\Variables=\emptyset$, we write $\GSTerms$ rather than $\STermsOn{\SSymbols}{\emptyset}$ for the set of \emph{ground} terms.
The set $\STerms$ of \emph{many-sorted terms} is $\STerms=\bigcup_{s\in S}\STerms_s$.
For $w=s_1\cdots s_n\in S^+$, we write $\STerms_w$ rather than $\STerms_{s_1}\times\cdots\times\STerms_{s_n}$ and even
write $\vec{t}\in\STerms_w$ rather than $t_i\in\STerms_{s_i}$ for each $1\leq i\leq n$.
The formulas $\varphi\in\SPformulas$ of a signature $\SPSignature$
are built up from atoms $P(\vec{t})$ with $P\in\Pi_{w}$ and $\vec{t}\in\STerms_{w}$, 
logic connectives ($\neg$, $\wedge$, and also $\vee$, $\Rightarrow$,...)
and quantifiers ($\forall$ and $\exists$) in the usual way.
A closed formula, i.e., one whose variables are all universally or existentially quantified, is called a \emph{sentence}.
In the following, substitutions $\sigma$ are assumed to be $S$-sorted mappings such that for all sorts $s\in S$, we have
$\sigma(x)\in\STerms_s$.

An $\SPSignature$-\emph{structure} $\SStructure$
consists of (i) a family $\{\SemDomain_s\mid s\in S\}$ of sets called the \emph{carriers}
or \emph{domains}
together with 
(ii) a function $f^\SStructure_{w,s}\in\SemDomain_w\to\SemDomain_s$ for each $f\in\SSymbols_{w,s}$
($\SemDomain_w$ is a one point set when $w=\lambda$ and hence $\SemDomain_w\to\SemDomain_s$ is isomorphic
to $\SemDomain_s$),
and 
(iii) an assignment
to each $P\in\Pi_w$ of a subset $P^\SStructure_w\subseteq\SStructure_{w}$; 
if the identity predicate $\_=\_:ss$ is in $\Pi_{ss}$, then $(=)^{\StructDomain}_{s\:s}=\{(a,a)\mid a\in \StructDomain_s\}$,
i.e.,  $\_=\_:ss$ is interpreted as the identity on $\SemDomain_s$.

Let 
$\SStructure$ and $\SStructure'$ be
$\SPSignature$-structures.
An \emph{$\SPSignature$-homomorphism} $h:\SStructure\to\SStructure'$ is 
an $S$-sorted function $h=\{h_s:\SemDomain_s\to\SemDomain'_s\mid s\in S \}$ such that
for each $f\in\SSymbols_{w,s}$ and $P\in\SPredicates_w$  with $w=s_1,\ldots,s_k$, (i)
$h_s(f^\SAlgebra_{w,s}(a_1,\ldots,a_k))=f^{\SAlgebra'}_{w,s}(h_{s_1}(a_1),\ldots,h_{s_k}(a_k))$
and (ii) if $\vec{a}\in P^{\SStructure}_w$, then $h(\vec{a})\in P^{\SStructure'}_w$. 
Given an $S$-sorted  \emph{valuation mapping} $\alpha:\Variables\to\SemDomain$, the evaluation mapping
$[\_]^\alpha_\SStructure:\STerms\to \SemDomain$ 
is the unique $(S,\SSymbols)$-homomorphism extending $\alpha$.
Finally, $[\_]^\alpha_\SStructure:\SPformulas\to\mathit{Bool}$ is given by: 
\begin{enumerate}
\item $[P(t_1,\ldots,t_n)]^\alpha_\SStructure=\mathit{true}$ (with $P\in\SPredicates_w$) 
if and only if $([t_1]^\alpha_\SStructure,\ldots,[t_n]^\alpha_\SStructure)\in P^\ModelDomain_w$;
\item $[\neg\varphi]^\alpha_\SStructure=\mathit{true}$ if and only if $[\varphi]^\alpha_\SStructure=\mathit{false}$;
\item $[\varphi\wedge\psi]^\alpha_\SStructure=\mathit{true}$ if and only if $[\varphi]^\alpha_\SStructure=\mathit{true}$ and $[\psi]^\alpha_\SStructure=\mathit{true}$; and
\item $[(\forall x:s)\:\varphi]^\alpha_\SStructure=\mathit{true}$ if and only if for all $a\in\SemDomain_s$, $[\varphi]^{\alpha[x\mapsto a]}_\SStructure=\mathit{true}$.
\end{enumerate}
A valuation $\alpha\in\Variables\to\SemDomain$ \emph{satisfies} $\varphi$ in  $\SStructure$ 
(written $\SStructure\models\varphi\:[\alpha]$) if $[\varphi]^\alpha_\SStructure=\mathit{true}$.
We then say that $\varphi$ is \emph{satisfiable}.
If $\SStructure\models\varphi\:[\alpha]$ for \emph{all} valuations $\alpha$, we write $\SStructure\models\varphi$
and say that $\SStructure$ is a \emph{model} of $\varphi$ or that $\varphi$ is \emph{true} in $\SStructure$.
We say that $\SModel$ is \emph{a model of a set of sentences} $\cS\subseteq\SPformulas$ (written $\SModel\models\cS$)
if for all $\varphi\in\cS$, $\SModel\models\varphi$.
Given a sentence $\varphi$, we write $\cS\models\varphi$ iff $\SModel\models\varphi$ holds for \emph{all models} 
$\SModel$ of $\cS$.

\section{Preservation of Many-Sorted First-Order Sentences}\label{SecPreservationManySorted}

Every set $\cS$ of ground atoms has an \emph{initial model} $\cI_\cS$ (or just $\cI$ if no confusion arises)
which consists of the usual (many-sorted) \emph{Herbrand Domain} of ground terms
modulo the equivalence $\sim$ generated by the equations in $\cS$. 
There is a unique homomorphism $h:\cI\to\SStructure$ from $\cI$ to any model 
$\SStructure$ of $\cS$ 
\cite[Section 3.2]{GogMes_ModelsAndEqualityForLogicalProgramming_TAPSOFT87}.
In the following, 
$h$ refers to such a homomorphism.
If $\cS$ contains no equation, then $\cI$ is  the (many-sorted) \emph{Least Herbrand Model} of $\cS$
and $\cI_s$ is $\GSTerms_s$ for each sort $s\in S$.
In the following, we consider sentences 
in \emph{prenex} form as follows:
\begin{eqnarray}
(Q_1 x_1:s_1)\cdots(Q_k x_k:s_k) \bigvee_{i=1}^m\bigwedge_{j=1}^{n_i}  L_{ij}\label{ManySortedClosureOfClauses}
\end{eqnarray} 
where 
(i) for all $1\leq i\leq m$ and $1\leq j\leq n_i$,
$L_{ij}$ are \emph{literals}, i.e., $L_{ij}=A_{ij}$ or $L_{ij}=\neg A_{ij}$ for some atom $A_{ij}$
(in the first case, we say that $L_{ij}$ is \emph{positive}; otherwise, it is \emph{negative}),
(ii) $x_1,\ldots,x_k$ for some $k\geq 0$ are the variables occurring in those literals (of sorts $s_1,\ldots,s_k$, respectively), and
(iii) $Q_1,\ldots,Q_k$ are universal/existential quantifiers.
A sentence $\varphi$ (equivalent to) (\ref{ManySortedClosureOfClauses}) is said to be \emph{positive} if all literals are.

\begin{theorem}
\label{TheoInitialModelAndValidityOfArbitraryFormulas}
Let $\SPSignature$ be a 
signature, 
$\cS$ be a set of ground atoms, 
$\varphi$ be a sentence (\ref{ManySortedClosureOfClauses}), 
and $\SStructure$ be a model of $\cS$
such that 
 (a) for all $q$, $1\leq q\leq k$,
 if $Q_q=\forall$ then $h_{s_q}$ 
 is surjective\footnote{A mapping $f:A\to B$ is \emph{surjective} if for all $b\in B$
there is $a\in A$ such that $f(a)=b$.} 
 and
  (b) for all negative literals $L_{ij}=\neg P(\vec{t})$, with $P\in\SPredicates_w$, 
and substitutions $\sigma$, 
if $h(\sigma(\vec{t}))\in P^\SStructure$ then $\sigma(\vec{t})\in P^\cI$.
 Then,
$\InitialModel_\cS\models\varphi \Longrightarrow \SStructure\models\varphi$.
\end{theorem}

\noindent
In order to achieve condition $(b)$ in Theorem \ref{TheoInitialModelAndValidityOfArbitraryFormulas},
given $P\in\SPredicates_w$, let $N(P)=\cI_w-P^\cI$ be the complement of the (Herbrand) 
interpretation of $P$.
Let $\cN(P)=\{\neg P(\vec{t})\mid \vec{t}\in N(P)\}$ (cf.\ Reiter's \emph{Closed World Assumption} \cite{Reiter_OnClosedWorldDatabases_1978}).
In general, $\cN(P)$ is infinite and incomputable.
In some simple cases, though, we can provide a \emph{finite description} of $\cN(P)$ for the required predicates $P$
(see Section \ref{SecDealingWithGeneralSentences}).
\begin{proposition}
\label{PropValidityOfArbitraryFormulasWithComplements}
Let $\SPSignature$ be a 
signature, 
 $\cS$ be a set of ground atoms, 
 $\varphi$ be a sentence (\ref{ManySortedClosureOfClauses}), 
$\SStructure$ be a model of $\cS$, 
and $\cN=\bigcup_{L_{ij}=\neg P(\vec{t})}\cN(P)$ be
 such that $\SStructure\models\cN$.
Let $L_{ij}=\neg P(\vec{t})$ be a  negative literal 
and $\sigma$ be a substitution.  
If $h(\sigma(\vec{t}))\in P^\SStructure$, then $\sigma(\vec{t})\in P^\cI$.
\end{proposition}
Consider a theory $\cS$ 
and let $\cS^\vdash$ be the 
set of ground atoms obtained as the \emph{deductive closure}  
of $\cS$, i.e., 
the set of \emph{atoms} 
$P(t_1,\ldots,t_n)$
for each $n$-ary predicate symbol $P$ and ground terms $t_1,\ldots,t_n$,  
such that $\cS\vdash P(t_1,\ldots,t_n)$.
The following result 
is the basis of the
practical applications discussed in the following sections.

\begin{corollary}[Semantic criterion]\label{CoroProofsAsSatisfiabilityPreservation}
Let $\SPSignature$ be a 
signature, 
 $\cS_0$ be a theory, 
  $\cS=\cS^\vdash_0$, 
 $\varphi$ be a sentence (\ref{ManySortedClosureOfClauses}), 
 and $\SStructure$ be a model of $\cS_0$
 such that 
 (a) for all $q$, $1\leq q\leq k$,
 if $Q_q=\forall$ then $h_{s_q}$ 
 is surjective
 and
  (b) for all negative literals $L_{ij}=\neg P(\vec{t})$, with $P\in\SPredicates_w$
and substitutions $\sigma$,  
if $h(\sigma(\vec{t}))\in P^\SStructure$ then $\sigma(\vec{t})\in P^\cI$.
If $\SStructure\models\neg\varphi$, then $\InitialModel_\cS\models\neg\varphi$.
\end{corollary}
In the following, we will not distinguish between theories $\cS$ and their ground deductive closure $\cS^\vdash$;
we rather use $\cS$ in both cases.
\begin{remark}[Proofs by satisfiability]
\label{ProofsBySemanticRefutation}
We can prove an arbitrary sentence $\varphi$ 
valid in $\cI_\cS$ by \emph{satisfiability} in some model $\SStructure$ of $\cS$. 
First define $\ol{\varphi}$ as the negation $\neg\varphi$ of $\varphi$. Then, 
find an appropriate structure $\SStructure$ satisfying $(a)$ and $(b)$
(with regard to $\ol{\varphi}$) and such that 
$\SStructure\models\cS\cup\{\neg\ol{\varphi}\}$.
By Corollary \ref{CoroProofsAsSatisfiabilityPreservation}, $\cI\models\neg\ol{\varphi}$ holds.
Since $\neg\ol{\varphi}$ is equivalent to $\varphi$, $\cI\models\varphi$ holds.
\end{remark}
Models $\SStructure$ to be used in Corollary \ref{CoroProofsAsSatisfiabilityPreservation} can be automatically generated from
the theory $\cS$ and sentence $\varphi$  by using 
a tool like \AGES\ \cite{GutLucRei_AToolForTheAutomaticGenerationOfLogicalModelsForOrderSortedFirstOrderTheories_PROLE16}
or Mace4.
In the following section, we investigate how to ensure \emph{surjectivity} when required in Corollary 
\ref{CoroProofsAsSatisfiabilityPreservation}.

\section{Surjective Homomorphisms}\label{SecSurjectiveHomomorphisms}

Given $\SPSignature=(S,\SSignature,\SPredicates)$, 
$s\in S$ and $T\subseteq\GSTerms_s$, consider the following sentences:\\[-0.4cm]
\begin{tabular}{c@{\hspace{-0.6cm}}c}
\begin{minipage}[t]{.515\linewidth}
\begin{eqnarray}
(\forall x:s)\bigvee_{t\in T} x=t\label{ReinforcingSurjectivityOfHomomorphismsViaGroundTerms}
\end{eqnarray}\nonumber
\end{minipage} & 
\begin{minipage}[t]{.52\linewidth}
\begin{eqnarray}
\bigwedge_{t,u\in T, t\neq u}\neg (t=u)\label{ReinforcingLowerBoundSizeOfSemanticDomain}
\end{eqnarray}\nonumber
\end{minipage}
\end{tabular}\\[0.2cm]
In the following, we write $(\ref{ReinforcingSurjectivityOfHomomorphismsViaGroundTerms})_s$ to make sort $s$ referred in (\ref{ReinforcingSurjectivityOfHomomorphismsViaGroundTerms}) explicit. We do the same in similar formulas below.
\begin{proposition}\label{PropReinforcingSurjectivityForFiniteStructures}
Let $\SPSignature$ 
be a signature, 
$\cS$ be a theory, 
$\SStructure$ be a model of $\cS$,
$s\in S$, 
and $T\subseteq\GSTerms_s$. 
(a) If $T\neq\emptyset$ and $\SStructure\models(\ref{ReinforcingSurjectivityOfHomomorphismsViaGroundTerms})_s$, then
 $h_s$ 
 is surjective and $|\SStructure_s|\leq|T|$.
(b) If $\SStructure_s\neq\emptyset$ and $\SStructure\models(\ref{ReinforcingLowerBoundSizeOfSemanticDomain})$, then
$|\SStructure_s|\geq|T|$.
\end{proposition}
In view of Proposition \ref{PropReinforcingSurjectivityForFiniteStructures}(a), denote 
$(\ref{ReinforcingSurjectivityOfHomomorphismsViaGroundTerms})_s$ as 
$\SurHomTh^T_s(\SPSignature)$ (or just $\SurHomTh^T_s$ or $\SurHomTh^T$ if no confusion arises).
Whenever $T$ is finite, Proposition \ref{PropReinforcingSurjectivityForFiniteStructures}(a) imposes that
the interpretation domain $\SStructure_s$ for sort $s$ is finite.
This is appropriate for tools like \MaceFour\ which generate structures with finite domains only.
However, the \emph{choice} of $T$ in Proposition \ref{PropReinforcingSurjectivityForFiniteStructures}, 
when used together with a theory $\cS$ imposing further requirements on symbols, 
can be crucial for Corollary \ref{CoroProofsAsSatisfiabilityPreservation} 
to succeed. 
Restricting the attention to finite domains can also be a drawback.
In the following, we investigate a different approach which avoids any choice of terms $T$ and
is valid for infinite structures as well.
Consider the following sentence:
\begin{eqnarray}
(\forall x:s)(\exists n:\SortNat)~
term_s(x,n)\label{ReinforcingSurjectivityOfHomomorphismsViaTerms}
\end{eqnarray}
where $\SortNat$ is a new sort, to be interpreted as the set $\naturals$ of natural numbers, and $term_s:s\:\SortNat$ is a new predicate
for each $s\in S$. 
The intended meaning of $(\ref{ReinforcingSurjectivityOfHomomorphismsViaTerms})_s$ is that, for all $x\in\SStructure_s$, 
there is $t\in\GSTerms_s$ of height at most  $n$ such that $x=t^\SStructure$.
We substantiate this, for each sort $s\in S$, 
by means of two (families of) formulas: 
\begin{eqnarray}
(\forall x:s)(\forall n:\SortNat)~term_s(x,0) & \Rightarrow & \bigvee_{c\in\SSymbols_{\lambda,s}} x=c\label{ReinforcingSurjectivityOfHomomorphismsViaTerms_termsBase}
\end{eqnarray}
{\footnotesize\begin{eqnarray}
(\forall x:s)(\forall n:\SortNat)(\exists m:\SortNat)~(n>0 \wedge term_s(x,n)) \Rightarrow\hspace{4cm} \nonumber\\
n>m \wedge \left ( term_s(x,m)\!\vee\!%
\bigvee_{\footnotesize\begin{array}{c}f\in\SSymbols_{w,s} \\ w\in S^+\end{array}}(\exists\vec{y}:w)\left (x=f(\vec{y})\wedge\bigwedge_{s_i\in w}terms_{s_i}(y_i,m)\right) \right )\hspace{0.3cm}
\label{ReinforcingSurjectivityOfHomomorphismsViaTerms_termsInductive}
\end{eqnarray}}%
Thus, by $(\ref{ReinforcingSurjectivityOfHomomorphismsViaTerms_termsBase})_s$, values $x$ satisfying $term_s(x,0)$
will be represented by some constant symbol $c$ of sort $s$.
Similarly, by $(\ref{ReinforcingSurjectivityOfHomomorphismsViaTerms_termsInductive})_s$, values $x$ satisfying $term_s(x,n)$
for some $n>0$ will be represented by some ground term $s$ of height $m$ for some $m<n$, 
or by a term $t=f(t_1,\ldots,t_k)$, where $f$ has rank $w\to s$ for some $w\in S^+$ and $t_1,\ldots,t_k$ have height $m$ at most.

The set $K(s)$ of $s$-\emph{relevant} sorts 
is 
the least set
satisfying: 
(i) $s\in K(s)$ and
(ii) if 
 $f\in\SSymbols_{s_1\cdots s_k,s'}$ and
$s'\in K(s)$, then $\{s_{1},\ldots,s_{n}\}\subseteq K(s)$.
Let
$\SPSignature_{\SortNat,s}=(S_\SortNat,\SSignature_\SortNat,\SPredicates_\mathit{\SortNat,K(s)})$ 
be an \emph{extension} of $\SPSignature$ where
$S_\SortNat=S\cup\{\SortNat\}$, 
$\SSignature_\SortNat$ extends $\SSignature$ with a new constant $\Fz:\lambda\to\SortNat$, 
and 
$\SPredicates_{\SortNat,K(s)}$ extends $\SPredicates$ with $>\::\SortNat\:\SortNat$
and a predicate $term_{s'}:s'\:\SortNat$ for each 
$s'\in K(s)$.
We let 
\begin{eqnarray}
\SurHomTh_s & = & \{(\ref{ReinforcingSurjectivityOfHomomorphismsViaTerms})_{s'},(\ref{ReinforcingSurjectivityOfHomomorphismsViaTerms_termsBase})_{s'},
(\ref{ReinforcingSurjectivityOfHomomorphismsViaTerms_termsInductive})_{s'}\mid s'\in K(s)\}
\end{eqnarray}
\begin{proposition}\label{PropSurjectiveHomomorphismWithOrderNat}
Let $\SPSignature$ 
be a signature, 
$\cS$ be a theory, 
$s\in S$, 
and $\SStructure$ be an $\SPSignature_{\SortNat,s}$-structure which is a model of $\cS$.
Assume that 
$\SStructure_\SortNat=\naturals$, 
$\Fz^\SStructure=0$,
and $\mt{m}>^\SStructure\mt{n}\Leftrightarrow \mt{m}>_\naturals \mt{n}$ for all $\mt{m},\mt{n}\in\SStructure_\SortNat$.
If $\SStructure\models\SurHomTh_s$, then $h_{s'}$ 
is surjective for all $s'\in K(s)$.
\end{proposition}
Given an extension $\SPSignature'$ of a signature $\SPSignature$, 
every $\SPSignature'$-structure $\SStructure'$ defines an $\SPSignature$-structure $\SStructure$: just take 
$\SStructure_s=\SStructure'_s$ for all $s\in S$, and then
$f^\SStructure_{w,s}=f^{\SStructure'}_{w,s}$ 
and 
$P^\SStructure_w=P^{\SStructure'}_{w}$ 
for all $w\in S^*$, $s\in S$, $f\in\SSymbols_{w,s}$,
and $P\in\SPredicates_w$.
Thus, Proposition \ref{PropSurjectiveHomomorphismWithOrderNat} is used to guarantee surjectivity of  
$h:\GSTerms_{s'}\to\SStructure_{s'}$,
rather than $h:\GSTermsOn{\SSignature_\SortNat}_{s'}\to\SStructure_{s'}$. 

\section{Examples of Application with Positive Sentences}\label{SecExampleOfApplication}

In this section we exemplify the use of Corollary \ref{CoroProofsAsSatisfiabilityPreservation} together with the approach in
Section \ref{SecSurjectiveHomomorphisms} to deal with
\emph{positive} sentences (\ref{ManySortedClosureOfClauses}), i.e., all literals are positive.

\subsection{Complete Definedness and Commutativity}

Consider the following \Maude\ specification 
(hopefully self-explained, but see \cite{ClavelEtAl_MaudeBook_2007}) 
for the arithmetic operations in Example \ref{ExAddMul} together with function 
$\pr{head}$, which returns the head of a list of natural numbers:
\begin{verbatim}
mod ExAddMulHead is
  sorts N LN . *** Sorts for numbers and lists of numbers
  op Z : -> N .        op suc : N -> N .  ops add mul : N N -> N .
  op head : LN -> N .  op nil : -> LN .   op cons : N LN -> LN .
  vars x y : N .       var xs : LN .
  rl add(Z,x) => x .   rl add(suc(x),y) => suc(add(x,y)) . 
  rl mul(Z,x) => Z .   rl mul(suc(x),y) => add(y,mul(x,y)) .
  rl head(cons(x,xs)) => x .
endm
\end{verbatim}
\emph{1) Complete definedness.}
We  claim \verb$head$ to be completely defined as follows:
\begin{eqnarray}
(\forall xs:\pr{LN})(\exists x:\pr{N})~\pr{head}(xs)\to x\label{ExHeadCompletelyDefined}
\end{eqnarray}
We \emph{disprove} (\ref{ExHeadCompletelyDefined}) by using Corollary \ref{CoroProofsAsSatisfiabilityPreservation}.
Due to the universal quantification of $xs$ in (\ref{ExHeadCompletelyDefined}), 
we need to ensure that 
$h_\pr{LN}:\GSTerms_\pr{LN}\to\SStructure_\pr{LN}$ 
is surjective for any structure $\SStructure$ we may use.
We use Proposition \ref{PropSurjectiveHomomorphismWithOrderNat}.
Since $K(\pr{LN})=\{\pr{N},\pr{LN}\}$ due to $\pr{cons}$, whose first argument is of sort $\pr{N}$,
$\SurHomTh_\pr{LN}$ consists of the following sentences:
\[
\begin{array}{l@{\hspace{1cm}}c}
(\forall x:\pr{N}) (\exists n:\SortNat)~ term_\pr{N}(x,n) & (\ref{ReinforcingSurjectivityOfHomomorphismsViaTerms})_{\pr{N}}\\
(\forall x:\pr{N})~term_\pr{N}(x,0)  \Rightarrow  x=\pr{Z} & (\ref{ReinforcingSurjectivityOfHomomorphismsViaTerms_termsBase})_{\pr{N}}\\
(\forall x:\pr{N})(\forall n:\SortNat)(\exists m:\SortNat)(\exists y:\pr{N})(\exists z:\pr{N}) (\exists ys:\pr{LN}) & (\ref{ReinforcingSurjectivityOfHomomorphismsViaTerms_termsInductive})_{\pr{N}}\\
\hspace{0.3cm}n > 0 \wedge term_\pr{N}(x,n) \Rightarrow  n > m \wedge [term_\pr{N}(x,m)~ \vee \\
\hspace{0.6cm} (term_\pr{N}(y,m) \wedge term_\pr{N}(z,m) \wedge term_\pr{LN}(ys,m)~\wedge \\
\hspace{0.9cm} (x=\pr{suc}(y)  \vee x=\pr{add}(y,z) \vee x=\pr{mul}(y,z) \vee x=\pr{head}(ys) ) ) ]\\[0.2cm]
(\forall xs:\pr{LN}) (\exists n:\SortNat)~ term_\pr{LN}(xs,n) & (\ref{ReinforcingSurjectivityOfHomomorphismsViaTerms})_{\pr{LN}}\\
(\forall xs:\pr{LN})~term_\pr{LN}(xs,0)  \Rightarrow  xs=\pr{nil} & (\ref{ReinforcingSurjectivityOfHomomorphismsViaTerms_termsBase})_{\pr{LN}}\\
(\forall xs:\pr{LN})(\forall n:\SortNat)(\exists m:\SortNat)(\exists y:\pr{N}) (\exists ys:\pr{LN})& (\ref{ReinforcingSurjectivityOfHomomorphismsViaTerms_termsInductive})_{\pr{LN}}\\
\hspace{0.3cm}n > 0 \wedge term_\pr{N}(x,n) \Rightarrow  n > m \wedge [term_\pr{N}(x,m)~ \vee \\
\hspace{0.6cm} (term_\pr{N}(y,m) \wedge term_\pr{LN}(ys,m)\wedge 
xs=\pr{cons}(y,ys)  ) ]
\end{array}
\]
We obtain a model $\SStructure$ of $\ol{\pr{ExAddMulHead}}\cup\SurHomTh_\pr{LN}\cup\{\neg(\ref{ExHeadCompletelyDefined})\}$ with 
\AGES. Sorts are interpreted as follows: $\SStructure_\pr{N}=\SStructure_\pr{LN}=\{-1,0\}$ 
and $\SStructure_\SortNat=\naturals$.
For function symbols:
\[\begin{array}{c@{\hspace{0.3cm}}c@{\hspace{0.3cm}}c@{\hspace{0.3cm}}c@{\hspace{0.3cm}}c}
\pr{Z}^\SStructure = -1 & \pr{nil}^\SStructure = 0 & \pr{suc}^\SStructure(x)=x & \pr{add}^\SStructure(x,y)=0\\ 
\pr{mul}^\SStructure(x,y)=0 & \pr{cons}^\SStructure(x,xs)=-1 & \pr{head}^\SStructure(xs) = -xs-1
\end{array}
\]
For predicates,
$x\to^\SStructure_\pr{N} y \Leftrightarrow x\geq y\wedge x\geq 0$,
$x\to^\SStructure_\pr{LN} y \Leftrightarrow x=y=-1$, and
both $x(\to^*_\pr{N})^\SStructure y$ and $x(\to^*_\pr{LN})^\SStructure y$ are \emph{true}.
We can check surjectivity of $h_s:\GSTerms_s\to\SStructure_s$ 
(for $s\in\{\pr{N},\pr{LN}\}$). For instance, we have:
\[
\begin{array}{rcl@{\hspace{0.5cm}}rcl@{\hspace{0.5cm}}l}
{}[\pr{Z}]_\SStructure & = & -1 & [\pr{add(Z,Z)}]_\SStructure & = & 0 & \text{for sort }\pr{N}\\
{}[\pr{cons(Z,nil)}]_\SStructure & = & -1 &  [\pr{nil}]_\SStructure & = & 0 & \text{for sort }\pr{LN}\\
\end{array}
\]

\smallskip
\noindent
\emph{2) Commutativity.} 
It is well-known that both $\Fadd$ and $\Fmul$ as defined by the rules of $\cR$ in Example \ref{ExAddMul} 
are \emph{commutative} on ground terms, i.e., for all ground terms $s$ and $t$,
$\Fadd(s,t)=_\cR\Fadd(t,s)$ and $\Fmul(s,t)=_\cR\Fmul(t,s)$, where $=_\cR$ is the equational theory induced by the rules $\ell\to r$ 
in $\cR$  treated as equations $\ell=r$.
Actually, by using Birkhoff's theorem and the fact that $\cR$ is \emph{confluent}, we can rephrase 
commutativity of $\Fadd$ as \emph{joinability} as follows:
\begin{eqnarray}
(\forall x)(\forall y) (\exists z)~ \Fadd(x,y)\to^*z\wedge \Fadd(y,x)\to^*z \label{PropCommutativityAddAsJoinability}
\end{eqnarray}
\begin{remark}
Proving commutativity of $\Fadd$ and $\Fmul$ when defined by $\cR$ in Example \ref{ExAddMul} by using Corollary 
\ref{CoroProofsAsSatisfiabilityPreservation} is possible (see Remark \ref{ProofsBySemanticRefutation}) but unlikely.
We should first define $\ol{\varphi}$ as $\neg(\ref{PropCommutativityAddAsJoinability})$, i.e., $\ol{\varphi}$ is
\begin{eqnarray}
(\exists x)(\exists y) (\forall z)~ \neg(\Fadd(x,y)\to^*z)\vee \neg(\Fadd(y,x)\to^*z) \label{PropCommutativityAddAsJoinabilityNegated}
\end{eqnarray}
Since (\ref{PropCommutativityAddAsJoinabilityNegated}) contains two negative literals, 
Corollary \ref{CoroProofsAsSatisfiabilityPreservation} requires the use of $\cN(\to^*)$.
\end{remark}
Since $\pr{head}$ is not completely defined, \verb$add$ and \verb$mul$ are \emph{not} commutative
in $\pr{ExAddMulHead}$.
We prove this fact by disproving the sorted version of (\ref{PropCommutativityAddAsJoinability}), i.e.,
\begin{eqnarray}
(\forall x:\pr{N})(\forall y:\pr{N}) (\exists z:\pr{N})~ \pr{add}(x,y)\to^*z\wedge \pr{add}(y,x)\to^*z \label{PropCommutativityAddAsJoinabilitySortN}
\end{eqnarray}
Due to the universal quantification of $x$ and $y$ in (\ref{PropCommutativityAddAsJoinabilitySortN}), 
we need to ensure that 
$h_\pr{N}:\GSTerms_\pr{N}\to\SStructure_\pr{N}$ 
is surjective.
Since $K(\pr{N})=\{\pr{N},\pr{LN}\}$ due to $\pr{head}$, 
we have
$\SurHomTh_\pr{N}=\SurHomTh_{\pr{LN}}$ as above.
\AGES\ obtain a model  $\SStructure$ of $\ol{\pr{ExAddMulHead}}\cup\SurHomTh_\pr{N}\cup\{\neg(\ref{PropCommutativityAddAsJoinabilitySortN})\}$
as follows: $\SStructure_\pr{N}=\{0,1\}$, $\SStructure_\pr{LN}=\{-1,0\}$ and $\SStructure_\SortNat=\naturals$.
Also,
\[\begin{array}{c@{\hspace{0.15cm}}c@{\hspace{0.2cm}}c@{\hspace{0.15cm}}c@{\hspace{0.2cm}}c}
\pr{Z}^\SStructure = 1 & \Fnil^\SStructure = -1 & \pr{suc}^\SStructure(x)=x & \pr{add}^\SStructure(x,y)=y\\ 
\pr{mul}^\SStructure(x,y)=x & \pr{cons}^\SStructure(x,xs)=x-1 & \pr{head}^\SStructure(xs) = xs+1 \\
x\to^\SStructure_\pr{N} y \Leftrightarrow x=y & x(\to^*_\pr{N})^\SStructure y \Leftrightarrow x=y &
x\to^\SStructure_\pr{LN} y \Leftrightarrow x=y & x(\to^*_\pr{LN})^\SStructure y \Leftrightarrow true
\end{array}
\]

\subsection{Top-Termination}\label{SecTopTermination}

A TRS $\cR$ is {\em top-terminating} if no infinitary reduction sequence performs 
infinitely many rewrites at topmost position $\toppos$ \cite{DerKapPla_RewriteRewriteRewriteRewrite_TCS91}.
From a computational point of view, top-termination is important in the semantic description of lazy languages
as it is an important ingredient to guarantee that every initial expression  has an infinite normal form \cite{DerKapPla_RewriteRewriteRewriteRewrite_TCS91,EndHen_LazyProductivityViaTermination_TCS11}.
Accordingly, given a \emph{dummy} sort $\genSort$, the \emph{negation} of 
\begin{eqnarray}
(\exists x:\genSort)(\forall n\in\naturals)(\exists y:\genSort)~x(\to^*\circ\exr)^ny\label{DefExistenceOfInfiniteTopReductionSequence}
\end{eqnarray}
(which claims for the existence of a term with infinitely many rewriting steps at top) 
captures top-termination. 
We introduce a new predicate $\to_{\star,\toppos}$ for the composition $\to^*\circ\exr$
of  the many-step rewriting relation $\to^*$  (defined as usual, i.e., by the whole theory $\ol{\cR}$ associated to
$\cR$) and topmost rewriting
$\exr$ defined by a theory $\cR_\toppos=\{(\forall\vec{x}:\genSort)~\ell \exr r\mid \ell\to r\in\cR\}$.
Sequences $s\to^{n}_{\star,\toppos}t$ meaning that $s$ $\to_{\star,\toppos}$-reduces into $t$ in $n+1$ $\to_{\star,\toppos}$-steps are defined as follows:
\begin{eqnarray}
(\forall x,y,z:\genSort) && x \to^*y\wedge y\exr z \Rightarrow x \to^0_{\star,\toppos} z\label{DefNReductionBaseCase_insertion}\\
(\forall x,y,z:\genSort) (\forall n\in\naturals) && x\to^0_{\star,\toppos} y \wedge y\to^n_{\star,\toppos} z\Rightarrow x  \to^{n+1}_{\star,\toppos} z\label{DefNReductionInductiveCaseStarToppos}
\end{eqnarray}
Overall, the sentence $\varphi$ to be disproved is:
\begin{eqnarray}
(\exists x:\genSort)(\forall n:\SortNat)(\exists y:\genSort)~x\to_{\star,\toppos}^ny\label{DefExistenceOfInfiniteTopReductionSequenceWithNat}
\end{eqnarray}
\begin{remark}
We use $\naturals$ in (\ref{DefExistenceOfInfiniteTopReductionSequence}) but $\SortNat$ in 
(\ref{DefExistenceOfInfiniteTopReductionSequenceWithNat}). Indeed,  (\ref{DefExistenceOfInfiniteTopReductionSequence})
is \emph{not} a valid sentence because
$\naturals$ is not first-order axiomatizable, see, e.g. \cite[Section 2.2]{Hodges_ModelTheory_1993}.
This is consistent with the well-known fact that termination (or top-termination) can\emph{not} be encoded in first-order logic
\cite[Section 5.1.4]{Shapiro_FoundationsWithoutFoundationalismACaseForSecondOrderLogic_1991}. We can use (\ref{DefExistenceOfInfiniteTopReductionSequenceWithNat}) together with Corollary \ref{CoroProofsAsSatisfiabilityPreservation} 
provided that $\SortNat$ is interpreted as $\naturals$. This is possible with \AGES.
\end{remark}

\begin{example}\label{ExSec9_5_EH11_TopTermination}
Consider the following (nonterminating) TRS $\cR$ \cite [Section 9.5]{EndHen_LazyProductivityViaTermination_TCS11}:
\\[-0.5cm] 
\begin{tabular}{c@{\hspace{-0.47cm}}c}
\begin{minipage}[t]{.515\linewidth}
\begin{eqnarray}
\Fnon & \to & \Ff(\Fg,\Ff(\Fnon,\Fg))\label{ExSec9_5_EH11_rule1}\\
\Fg & \to & \Fa\label{ExSec9_5_EH11_rule2}
\end{eqnarray}\nonumber
\end{minipage} & 
\begin{minipage}[t]{.52\linewidth}
\begin{eqnarray}
\Ff(\Fa,x) & \to & \Fa\label{ExSec9_5_EH11_rule3}\\
\Ff(\Fb,\Fb) & \to & \Fb\label{ExSec9_5_EH11_rule4}\\
\Ff(\Fb,\Fa) & \to & \Fb\label{ExSec9_5_EH11_rule5}
\end{eqnarray}\nonumber
\end{minipage}
\end{tabular}\\[0.3cm]
The associated theory $\cR_\mathit{topT}$ is $\cR_\mathit{topT}=\ol{\cR}\cup\cR_\toppos\cup\{(\ref{DefNReductionBaseCase_insertion}),(\ref{DefNReductionInductiveCaseStarToppos})\}$, where
$\cR_\toppos$ is
\\[-0.5cm] 
\begin{tabular}{c@{\hspace{-0.6cm}}c}
\begin{minipage}[t]{.515\linewidth}
\begin{eqnarray}
 \Fnon & \exr & \Ff(\Fg,\Ff(\Fnon,\Fg))\label{ExSec9_5_EH11_TopRewriting_sentence1}\\
\Fg  & \exr &  \Fa\label{ExSec9_5_EH11_TopRewriting_sentence2}\\
\end{eqnarray}\nonumber
\end{minipage} & 
\begin{minipage}[t]{.52\linewidth}
\begin{eqnarray}
(\forall x:\genSort)~ \Ff(\Fb,x)  & \exr &  \Fb\label{ExSec9_5_EH11_TopRewriting_sentence3}\\
 \Ff(\Fb,\Fb)   & \exr &  \Fb\label{ExSec9_5_EH11_TopRewriting_sentence4}\\
 \Ff(\Fb,\Fa)   & \exr &  \Fb\label{ExSec9_5_EH11_TopRewriting_sentence5}
\end{eqnarray}\nonumber
\end{minipage}
\end{tabular}\\[0.3cm]
Note that (\ref{DefExistenceOfInfiniteTopReductionSequenceWithNat}) only requires that the homomorphism mapping terms of sort
$\SortNat$ to $\nat$ is surjective, which is automatically achieved by \AGES.
The structure $\SStructure$ with $\SStructure_\genSort=\{-1,0,1\}$, $\SStructure_{\SortNat}=\naturals$, function symbols 
interpreted by:
$\Fa^\SStructure  =  1$,
$\Fb^\SStructure =  1$,
$\Fg^\SStructure =  0$,
$\Fnon^\SStructure = -1$, and
$\Ff^\SStructure(x) = 0$;
and predicate symbols as follows:
\[\begin{array}{r@{\:}c@{\:}l@{\hspace{0.6cm}}r@{\:}c@{\:}l@{\hspace{0.6cm}}r@{\:}c@{\:}l@{\hspace{0.6cm}}r@{\:}c@{\:}l}
 x \to^\SStructure y & \Leftrightarrow & y\geq x\wedge x+y\geq -1 &  x (\to^*)^\SStructure y & \Leftrightarrow & y\geq x \\
  x (\exr)^\SStructure y & \Leftrightarrow & y> x & x (\to^n_{\star,\toppos})^\SStructure y & \Leftrightarrow & y>x+ n
 \end{array}
\]
is a model of $\cR_{\mathit{topT}}\cup\{\neg (\ref{DefExistenceOfInfiniteTopReductionSequenceWithNat})\}$ and proves top-termination of 
$\cR$.
\end{example}

\section{Refutation Witnesses}\label{SecWitnessesOfSemanticRefutation}

In logic, a \emph{witness} for an existentially quantified sentence $(\exists x)\varphi(x)$ is a specific value $b$ to be substituted
by $x$ in $\varphi(x)$ so that $\varphi(b)$ is true (see, e.g., \cite[page 81]{BooBurJef_ComputabilityAndLogic_2002}).
Similarly, we can think of a value $b$ such that $\neg\varphi(b)$ holds as a witness of $(\exists x)\neg\varphi(x)$ or as
a \emph{refutation witness} for $(\forall x)\varphi(x)$;
we can also think of $b$ as a \emph{counterexample} to $(\forall x)\varphi(x)$ 
\cite[page 284]{Kleene_MathematicalLogic_1967}.
Note, however, that 
witnesses that are given as values $b$ belonging to an \emph{interpretation domain} $\SStructure$ can be
meaningless for the user who is acquainted with the first-order language $\SPSignature$ but not so much with abstract values
from $\SStructure$ (which is often automatically synthesized by using some tool).
Users can be happier to deal with \emph{terms} $t$ which are somehow connected to witnesses $b$ by a homomorphism, so that 
$t^\SStructure=b$.
Corollary \ref{CoroProofsAsSatisfiabilityPreservation} permits a refutation of $\varphi$ 
by finding a model $\SStructure$ of $\neg\varphi$ to conclude that $\cI\models\neg\varphi$. We want to obtain 
instances of $\varphi$ 
to better understand unsatisfiability of
$\varphi$.
In this section we investigate this problem.

The negation $\neg(\ref{ManySortedClosureOfClauses})$ of (\ref{ManySortedClosureOfClauses}), i.e., of
$(Q_1 x_1:s_1)\cdots(Q_k x_k:s_k) \bigvee_{i=1}^m\bigwedge_{j=1}^{n_i}  L_{ij}$ is
\begin{eqnarray}
(\ol{Q}_1 x_1:s_1)\cdots(\ol{Q}_k x_k:s_k) \bigwedge_{i=1}^m\bigvee_{j=1}^{n_i}  \neg L_{ij}(x_1,\ldots,x_k)\label{ManySortedClosureOfClausesNegation}
\end{eqnarray}
where $\ol{Q}_i$ is $\forall$ whenever $Q_i$ is $\exists$ and $\ol{Q}_i$ is $\exists$ whenever $Q_i$ is $\forall$.
We assume 
$\eta\leq k$ universal quantifiers in (\ref{ManySortedClosureOfClausesNegation}) with indices 
$U=\{\upsilon_1,\ldots,\upsilon_\eta\}\subseteq\{1,\ldots,k\}$ 
and hence $k-\eta$ existential quantifiers with indices $E=\{\epsilon_1,\ldots,\epsilon_{k-\eta}\}=\{1,\ldots,k\}-U$.
In the following $\ol{\eta}$ denotes $k-\eta$.
For each $\epsilon\in E$, we let $U_\epsilon=\{\upsilon\in U\mid \upsilon<\epsilon\}$ be the (possibly empty) set of 
indices of universally quantified variables in (\ref{ManySortedClosureOfClausesNegation}) occurring before $x_\epsilon$ in
the quantification prefix of (\ref{ManySortedClosureOfClausesNegation}). 
Let $\eta_\epsilon=|U_\epsilon|$.
Note that $U_{\epsilon_1}\subseteq U_{\epsilon_2}\subseteq\cdots \subseteq U_{\epsilon_{\ol{\eta}}}$.
Let $U_{\exists}$ be the set  of indices of universally quantified variables occurring \emph{before}
some existentially quantified variable in the quantification prefix of (\ref{ManySortedClosureOfClausesNegation}).
Note that $U_{\exists}$ is empty whenever $\upsilon_1>\epsilon_{k-\eta}$  (no existential quantification after a universal quantification);
otherwise, $U_{\exists}=\{\upsilon_1,\ldots,\upsilon_{\exists}\}$ for some $\upsilon_{\exists}\leq\upsilon_\eta$.
Accordingly, $U_\forall=U-U_\exists=\{\epsilon_{\ol{\eta}}+1,\ldots,k\}$ 
 is the set of indices of universally quantified variables occurring \emph{after}
all existentially quantified variables in the quantification prefix of (\ref{ManySortedClosureOfClausesNegation}).
Note that $U_\forall$ is empty whenever $\epsilon_1>\upsilon_\eta$  (no universal quantification after an existential quantification).

Most theorem provers transform sentences into universally quantified formulas by Skolemization (see, e.g., \cite{KimZha_ModGenTheoremProvingByModelGeneration_AAAI94}). Thus, if $k>\eta$, i.e., (\ref{ManySortedClosureOfClausesNegation}) contains
existential quantifiers, 
we need to introduce \emph{Skolem function symbols} $sk_\epsilon:w_\epsilon\to s_\epsilon$ for each 
$\epsilon\in E$,
where $w_\epsilon$ is the (possibly empty) sequence of $\eta_\epsilon$ sorts indexed by $U_\epsilon$. 
Note that $sk_\epsilon$ is a \emph{constant} if $\eta_\epsilon=0$.
The \emph{Skolem normal form} of (\ref{ManySortedClosureOfClausesNegation}) is
\begin{eqnarray}
(\forall x_{\upsilon_1}:s_{\upsilon_1})\cdots(\forall x_{\upsilon_\eta}:s_{\upsilon_\eta})\bigwedge_{i=1}^m\bigvee_{j=1}^{n_i}  \neg L_{ij}(e_1,\ldots,e_k)\label{ManySortedClosureOfClausesNegationSkolemNormalForm}
\end{eqnarray}
where for all $1\leq q\leq k$,
(i) $e_q\equiv x_q$ if $q\in U$ and 
(ii) $e_q\equiv sk_q(\vec{x}_{\eta_q})$ if $q\in E$, where $\vec{x}_{\eta_q}$ is the sequence of
variables $x_{\nu_1},\ldots,x_{\nu_{\eta_q}}$.
If $E\neq\emptyset$ (i.e., (\ref{ManySortedClosureOfClausesNegation}) and 
(\ref{ManySortedClosureOfClausesNegationSkolemNormalForm}) differ), then
(\ref{ManySortedClosureOfClausesNegationSkolemNormalForm}) is a sentence of an \emph{extended} signature 
$\SPSignature^{sk}=(S,\SSignature^{sk},\SPredicates)$
where $\SSignature^{sk}$ extends $\SSignature$ with skolem functions.
Since (\ref{ManySortedClosureOfClausesNegationSkolemNormalForm})  logically implies 
(\ref{ManySortedClosureOfClausesNegation}) \cite[Section 19.2]{BooBurJef_ComputabilityAndLogic_2002},
every model $\SStructure$ of $(\ref{ManySortedClosureOfClausesNegationSkolemNormalForm})$
is a model of $(\ref{ManySortedClosureOfClausesNegation})$ as well.

\begin{definition}[Set of refutation witnesses]
\label{DefSetOfRefutationWitnesses}
Using the notation developed in the previous paragraphs, let 
$\SStructure$ be an  $\SPSignature^{sk}$-structure such that 
$h_{s_q}$ 
is \emph{surjective} for all $q\in U_\exists\cup E$.
The $\SPSignature^{sk}$-sentence (\ref{ManySortedClosureOfClausesNegationSkolemNormalForm}) is given a \emph{set 
of refutation witnesses} $\Phi$ consisting of 
 $\SPSignature$-sentences $\phi_\alpha$
for each valuation $\alpha$ of the variables $x_{\upsilon_1},\ldots,x_{\upsilon_{\exists}}$ indexed by $U_\exists$; 
each $\phi_\alpha$
is (nondeterministically) defined as follows: 
\begin{eqnarray}
(\forall x_{\epsilon_{\ol{\eta}}+1}:s_{\epsilon_{\ol{\eta}}+1})\cdots(\forall x_k:s_k)\bigwedge_{i=1}^m\bigvee_{j=1}^{n_i}  \neg L_{ij}(e'_1,\ldots,e'_k)\label{ManySortedClosureOfClausesNegationInterpreted}
\end{eqnarray}
where for all $1\leq q\leq k$,
(i) $e'_q\equiv x_q$ if $q\in U_\forall$ and
(ii)  $e'_q\equiv t$ if $q\in U_\exists\cup E$ and $t\in\GSTerms_{s_q}$ is such that $[t]_\SStructure=[e_q]^\alpha_\SStructure$.
\end{definition}
Note that, in Definition \ref{DefSetOfRefutationWitnesses} we could 
emph{fail} to find the necessary terms $t\in\GSTerms_{s_q}$ if $h_{s_q}$ is \emph{not} surjective.
Note also that, whenever $E$ is empty, $\Phi$ is a singleton consisting of 
(\ref{ManySortedClosureOfClausesNegationInterpreted}) which coincides with (\ref{ManySortedClosureOfClausesNegationSkolemNormalForm}).
We have the following:

\begin{proposition}\label{PropRefutationWitnesses}
For every  $\SPSignature^{sk}$-structure $\SStructure$,
$\SStructure\models(\ref{ManySortedClosureOfClausesNegationSkolemNormalForm})$ if and only if
$\SStructure\models \Phi$.
\end{proposition}
Refutation witnesses are built
from symbols in the original signature $\SPSignature$ only. 
We can use them as more intuitive \emph{counterexamples} to the refuted property $\varphi$.

\begin{proposition}\label{PropRefutationWitnessesHoldsInCanonicalModel}
Let $\SPSignature$ be a 
signature, 
$\cS$ be a theory, 
$\varphi$ be a sentence (\ref{ManySortedClosureOfClauses}), 
and $\SStructure$ be a model of $\cS$ such that 
for all negative literals $L_{ij}=\neg P(\vec{t})$ 
with $P\in\SPredicates_w$
and substitutions $\sigma$,
if $h(\sigma(\vec{t}))\in P^\SStructure$ then
$\sigma(\vec{t})\in P^\cI$. For all 
$\phi\in\Phi$, $\cI\models \phi$.
\end{proposition}

\begin{corollary}\label{CoroPositiveRefutationWitnessesHoldsInCanonicalModel}
If (\ref{ManySortedClosureOfClauses}) is positive, then for all refutation witnesses $\phi\in\Phi$, $\cI\models \phi$.
\end{corollary}

\begin{example}
Consider $\pr{ExAddMulHead}$ in Section \ref{SecExampleOfApplication}. 
The \emph{refutation} of (\ref{ExHeadCompletelyDefined}) using \AGES\
actually proceeds by skolemization of the negation of (\ref{ExHeadCompletelyDefined}), i.e., of
\begin{eqnarray}
(\exists xs:\pr{LN})(\forall x:\pr{N})~\neg(\pr{head}(xs)\to x)~~~\label{ExHeadCompletelyDefined_Negation}
\end{eqnarray}
With regard to (\ref{ExHeadCompletelyDefined_Negation}), we have
$E=\{1\}$, $U_\exists=\emptyset$ and $U_\forall=\{2\}$, where $1$ and $2$ refer to variables $xs$ and $x$,
respectively.
Accordingly, $\upsilon_\exists=0$.
The only sort involved in the variables indexed by $U_\exists\cup E$ is $\pr{LN}$. 
Since variables of sort $\pr{LN}$ are universally quantified in (\ref{ExHeadCompletelyDefined}),
the application of Corollary \ref{CoroProofsAsSatisfiabilityPreservation} 
in  Section \ref{SecExampleOfApplication} already required surjectivity of $h_\pr{LN}$.
The \emph{Skolem normal form} of (\ref{ExHeadCompletelyDefined_Negation}) is:
\begin{eqnarray}
(\forall x:\pr{N})~\neg(\pr{head}(\Fsk_{xs})\to x)~~~\label{ExHeadCompletelyDefined_NegationSkolemNormalForm}
\end{eqnarray}
where $\Fsk_{xs}$ is a new constant of sort $\pr{LN}$.
The structure $\SStructure$ computed by \AGES\ 
is actually a model of 
$\ol{\cR}\cup\SurHomTh_\pr{LN}\cup\{(\ref{ExHeadCompletelyDefined_NegationSkolemNormalForm})\}$,
for $\SurHomTh_\pr{LN}$ in Section \ref{SecExampleOfApplication}.
For $\Fsk_{xs}$, we have $\Fsk^\SStructure_{xs}=0$.
There is a single (empty) valuation $\alpha$ of variables indexed by $U_\exists$ (which is empty). 
Hence, $\Phi=\{\phi_\alpha\}$ is a singleton.
According to Definition \ref{DefSetOfRefutationWitnesses}, since $[\pr{nil}]_\SStructure=0=[\Fsk_{xs}]_\SStructure$,
the following sentence could be associated to the refutation witness $\phi_\alpha$: 
$(\forall x:\pr{N})~\neg(\pr{head}(\pr{nil})\to x)$.
\end{example}

\begin{example}
With regard to the computation of refutation witnesses for $\cR$ in Example \ref{ExSec9_5_EH11_TopTermination}, 
we start with the negation of (\ref{DefExistenceOfInfiniteTopReductionSequence}), i.e.,
\begin{eqnarray}
(\forall x:\genSort)(\exists n:\SortNat)(\forall y:\genSort)~\neg(x(\to^*\circ\exr)^ny)\label{DefExistenceOfInfiniteTopReductionSequence_Negation}
\end{eqnarray}
We have $E=\{2\}$, $U_\exists=\{1\}$ and $U_\forall=\{3\}$. The Skolem normal form of (\ref{DefExistenceOfInfiniteTopReductionSequence_Negation}) is
\begin{eqnarray}
(\forall x:\genSort)(\forall y:\genSort)~\neg(x(\to^*\circ\exr)^{\Fsk_n(x)}y)\label{DefExistenceOfInfiniteTopReductionSequence_NegationSkolemNF}
\end{eqnarray}
where $\Fsk_n:\genSort\to\SortNat$ is a new (monadic) function symbol.
Since the sorts for variables indexed by $U_\exists\cup E$ are $\genSort$ and $\SortNat$, we require surjectivity of 
$h_\genSort$ and $h_\SortNat$.
This is achieved by using $\SurHomTh_\genSort$ and interpreting $\SortNat$ as $\naturals$ 
as done in \AGES.
The structure $\SStructure$ in Example \ref{ExSec9_5_EH11_TopTermination} is a model of 
$\cR_{\mathit{topT}}\cup\SurHomTh_\genSort\cup\{(\ref{DefExistenceOfInfiniteTopReductionSequence_NegationSkolemNF})\}$.
The interpretation obtained for $\Fsk_n$ is
\[\Fsk^\SStructure_n(x)=1-x\]
Now we can compute refutation witnesses for (\ref{DefExistenceOfInfiniteTopReductionSequence_NegationSkolemNF}).
Since $U_\epsilon=\{1\}$ is a singleton whose index refers to a variable $x$ of sort $\genSort$ and $\SStructure_\genSort=\{-1,0,1\}$, 
we have to deal with three valuation functions for the only variable $x$ to be considered:
\[\begin{array}{rcl@{\hspace{1cm}}rcl@{\hspace{1cm}}rcl}
\alpha_{-1}(x) & = & -1 & \alpha_0(x) & = & 0 & \alpha_1(x) & = & 1 
\end{array}
\]
We have $\Phi = \{\phi_{\alpha_{-1}},\phi_{\alpha_0},\phi_{\alpha_1}\}$, where 
$\phi_{\alpha_{-1}}$ is 
$(\forall y:\genSort)  \neg(\Fnon(\to^*\circ\exr)^2y)$, 
$\phi_{\alpha_{0}}$ is 
$(\forall y:\genSort) \neg(\Fg(\to^*\circ\exr)^{1}y)$, and 
$\phi_{\alpha_{1}}$ is
$(\forall y:\genSort) \neg(\Fa(\to^*\circ\exr)^{0}y)$.
\end{example}
Note that, since $\Ff^\SStructure(x)=0$, we could also write $\phi_{\alpha_{0}}$ as
$(\forall y:\genSort)~\neg(\Ff(t)(\to^*\circ\exr)^{1}y)$
for \emph{every ground term $t$}. This gives additional, complementary information. 

\section{Example of Application with General Sentences}\label{SecDealingWithGeneralSentences}

Consider a well-known example of a \emph{locally confluent} but \emph{nonconfluent} TRS  $\cR$:
\[
\begin{array}{rcl@{\hspace{1cm}}rcl@{\hspace{1cm}}rcl@{\hspace{1cm}}rcl@{\hspace{1cm}}}
\Fb & \to & \Fa &
\Fb & \to & \Fc &
\Fc & \to & \Fb &
\Fc & \to & \Fd
\end{array}
\]

\begin{example}[Local confluence of $\cR$]\label{ExWCRnoCR_proofOfWCR}
Local confluence corresponds to $\varphi_{WCR}$ in 
Table \ref{TableSomeFOpropertiesRewritingBasedSystems}.
As explained in Remark \ref{ProofsBySemanticRefutation},
we start with 
$\ol{\varphi}_{\mathit{WCR}}=\neg\varphi_{\mathit{WCR}}$
i.e.,
\\[-0.8cm]

{\footnotesize\begin{eqnarray}
(\exists x,y,z:\genSort)(\forall u:\genSort)~(x\to y\wedge x\to z \wedge \neg(x\to^*u))\vee (x\to y\wedge x\to z \wedge \neg(z\to^*u))~~~
\end{eqnarray}}%
Due to the universal quantifier, 
$h_\genSort:\GSTerms_\genSort\to\SStructure_\genSort$ must be surjective.
We can achieve this by adding the following sentence $\SurHomTh^T_\genSort$ for $T=\{\Fa,\Fb,\Fc,\Fd\}$:
\begin{eqnarray}
(\forall x:\genSort)~x=\Fa\vee x=\Fb\vee x=\Fc\vee x=\Fd 
\end{eqnarray}
Due to the negative literals $\neg(x\to^*u)$ and $\neg(z\to^*u)$, we consider $\cN$, representing the 
forbidden many-step rewriting steps, explicitly given by:
\[\cN=\{
\begin{array}{c@{\hspace{0.1cm}}c@{\hspace{0.1cm}}c@{\hspace{0.1cm}}c@{\hspace{0.1cm}}c@{\hspace{0.1cm}}c@{\hspace{0.1cm}}c}
\neg(\Fa \to^* \Fb), & 
\neg(\Fa \to^* \Fc), & 
\neg(\Fa \to^* \Fd), & 
\neg(\Fd \to^* \Fa), & 
\neg(\Fd \to^* \Fb), & 
\neg(\Fd \to^* \Fc)
\end{array}
\}
\]
We apply Corollary \ref{CoroProofsAsSatisfiabilityPreservation} to prove that $\neg\ol{\varphi}_{\mathit{WCR}}$ (i.e., $\varphi_{\mathit{WCR}}$) holds 
by obtaining a model of $\ol{\cR}\cup\SurHomTh^T_\genSort\cup \cN\cup\{\varphi_{\mathit{WCR}}\}$ 
with \MaceFour.\footnote{This proves $\cR$ \emph{ground} locally confluent, i.e., variables in $\varphi_{\mathit{WCR}}$ 
refer to \emph{ground} terms only; since $\cR$ is a ground TRS, local confluence and ground local  confluence coincide.}
The structure has domain $\SStructure_\genSort=\{0,1,2,3\}$; constants are interpreted as follows:
$\Fa^\SStructure = 0$, 
$\Fb^\SStructure = 1$, 
$\Fc^\SStructure=3$, 
and $\Fd^\SStructure=2$.
With regard to predicate symbols, we have:
\[\begin{array}{c@{\hspace{0.5cm}}c@{\hspace{0.5cm}}c@{\hspace{0.5cm}}c@{\hspace{0.5cm}}c}
x\to^\SStructure y = \{(1,0),(1,3),(3,1),(3,2)\} & x(\to^*)^\SStructure y = \{(1,x),(3,x)\mid x\in\SStructure_\genSort\}
\end{array}
\]
This proves $\cR$ locally confluent.
\end{example}

\begin{example}[Nonconfluence of $\cR$]\label{ExWCRnoCR_disProofOfCR}
In order to \emph{disprove} confluence of $\cR$, which is represented by $\varphi_{CR}$ in 
Table \ref{TableSomeFOpropertiesRewritingBasedSystems}, we first write $\varphi_{CR}$
 in the form (\ref{ManySortedClosureOfClauses}), i.e.,
\begin{eqnarray}
(\forall x,y,z:\genSort)(\exists u:\genSort)~\neg(x\to^*y)\vee \neg(x\to^*z)\vee (y \to^*u\wedge z\to^*u)~
\end{eqnarray}
Due to the universal quantification and negative literals,
we use $\SurHomTh^T_\genSort$ and
$\cN$ as in Example \ref{ExWCRnoCR_proofOfWCR}.
We obtain a model $\SStructure$ of $\ol{\cR}\cup\SurHomTh^T_\genSort\cup \cN\cup\{\neg\varphi_{\mathit{CR}}\}$ with \MaceFour.
The domain is $\SemDomain_\genSort=\{0,1,2\}$ and symbols are interpreted by:
$\Fa^\SStructure = 0$, 
$\Fb^\SStructure=\Fc^\SStructure = 1$, 
$\Fd^\SStructure=2$, 
$x\to^\SStructure y \Leftrightarrow x = 1$, and
$x(\to^*)^\SStructure y \Leftrightarrow x = y\vee x=1$.
This proves nonconfluence of $\cR$. 
With regard to the refutation witnesses, 
$\neg\varphi_{CR}$ is
\begin{eqnarray}
(\exists x,y,z:\genSort)(\forall u:\genSort)~x\to^*y\wedge x\to^*z\wedge \neg(y \to^*u\wedge z\to^*u)~
\end{eqnarray}
and its Skolem normal form is
\begin{eqnarray}
(\forall u:\genSort)~\Fsk_x\to^*\Fsk_y\wedge \Fsk_x\to^*\Fsk_z\wedge \neg(\Fsk_y \to^*u\wedge \Fsk_z\to^*u)~
\end{eqnarray}
$\MaceFour$ yields 
$\Fsk_x^\SStructure=1$, $\Fsk_y^\SStructure=0$ and $\Fsk_z^\SStructure=2$;
$\Phi$ consists of a single sentence; e.g.,
\begin{eqnarray}
(\forall u:\genSort)~\Fb\to^*\Fa\wedge \Fb\to^*\Fd\wedge \neg(\Fa \to^*u\wedge \Fd\to^*u)~
\end{eqnarray}
but also:
$(\forall u:\genSort)~\Fc\to^*\Fa\wedge \Fc\to^*\Fd\wedge \neg(\Fa \to^*u\wedge \Fd\to^*u)$.
Indeed, they represent the two possible cases of nonconfluent behavior in $\cR$.
\end{example}

\begin{example}[Normalizing TRS]
$\cR$ is not terminating, but we can prove it \emph{normalizing} (i.e., every term has a normal form) by 
\emph{disproving} $\ol{\varphi}_{\mathit{WN}}$, 
for $\varphi_{\mathit{WN}}$ in Table \ref{TableSomeFOpropertiesRewritingBasedSystems}.
Therefore, $\ol{\varphi}_{\mathit{WN}}$ is
$(\exists x:\genSort)(\forall y:\genSort)(\exists z:\genSort)\:(\neg(x\to^* y)\vee y\to z)$.
We guarantee surjectivity by using $\SurHomTh^T_\genSort$ in Example \ref{ExWCRnoCR_proofOfWCR};
we also use $\cN$ in Example \ref{ExWCRnoCR_proofOfWCR}.
\MaceFour\ obtains  a model $\SStructure$ of $\ol{\cR}\cup\SurHomTh^T_\genSort\cup \cN\cup\{\varphi_{\mathit{WN}}\}$ 
with $\SemDomain_\genSort=\{0,1,2\}$, 
$\Fa^\SStructure = 0$,
$\Fb^\SStructure=\Fc^\SStructure = 1$, 
$\Fd^\SStructure=2$,
$x\to^\SStructure y \Leftrightarrow x = 1$, and 
$x(\to^*)^\SStructure y \Leftrightarrow x = y\vee x=1$.
\end{example}

\section{Related Work}\label{SecRelatedWork}

In \cite[Section 6]{Lucas_AnalysisOfRewritingBasedSystemsAsFirstOrderTheories_LNCS_LOPSTR17} 
we already compared our approach to
existing techniques and tools for the so-called First-Order Theory of Rewriting \cite{DauTis_TheTheoryOfGroundRewriteSystemsIsDecidable_LICS90},
which applies to restricted classes of TRSs and formulas.
In \cite{LucGut_UseOfLogicalModelsForProvingInfeasibilityInTermRewriting_IPL18},
we show that our semantic approach is practical when applied to arbitrary (Conditional) TRSs.

McCune's \ProverNine/\MaceFour\ 
are popular automated systems for theorem proving in first-order and equational 
logic. 
Given a theory $\cS$ and a \emph{goal} or \emph{statement} $\varphi$, 
\ProverNine\ tries to prove that 
$\cS\vdash\varphi$ holds. 
The generator of models \MaceFour\  
complements \ProverNine\ as follows: ``\emph{If the statement is the denial of some conjecture, any structures found by \MaceFour\ are counterexamples to the conjecture}''.\footnote{\url{https://www.cs.unm.edu/~mccune/prover9/manual/2009-11A/mace4.html}}
Accordingly,
the user introduces $\varphi$ in the \emph{goal} section of \MaceFour, but the system seeks a model of $\cS\cup\{\neg\varphi\}$.
Indeed, as discussed in Section \ref{SecIntroduction}, 
if $\SStructure\models\cS\cup\{\neg\varphi\}$ holds, then 
$\cS\vdash\varphi$ does not hold.
But, unless $\varphi$ is an ECBCA, this does not necessarily mean that $\varphi$ \emph{does not hold} of a program $P$
with $\cS=\ol{P}$!
Consider the following `misleading' session with \MaceFour\ that `\emph{disproves}' commutativity of the addition.

\begin{example}\label{ExAddMulCommutativity}
Consider $\cR$ 
in Example \ref{ExAddMul}.
\MaceFour\ obtains a model $\SStructure$ of $\ol{\cR}\cup\{\neg(\ref{PropCommutativityAddAsJoinability})\}$ with domain
$\SStructure=\{0,1\}$, and function and predicate symbols as follows:
$\Fz^\SStructure  =  0$, 
$\Fs^\SStructure(x)  =  x$, 
$\Fadd^\SStructure(x,y)  =  
\left \{
\begin{array}{cl}
1 & \text{if } x = 0 \wedge y=1\\
0 & \text{otherwise}
\end{array}
\right .$,
$\Fmul^\SStructure(x,y) = 0$,
and $\to^\SStructure$ and $(\to^*)^\SStructure$ both interpreted as the equality.
Additionally, \MaceFour\ also displays the following: $\verb!c1!^\SStructure = 0$ and  $\verb!c2!^\SStructure  = 1$.
These  \verb!c1! and \verb!c2! are \emph{new} Skolem symbols (but \emph{unexpected} for most users!). 
In practice, \MaceFour\ finds a model for the \emph{Skolem normal form} 
of $\neg(\ref{PropCommutativityAddAsJoinability})$, which is
\begin{eqnarray}
(\forall z)~ \neg(\Fadd(\verb!c1!,\verb!c2!)\to^*z\wedge \Fadd(\verb!c2!,\verb!c1!)\to^*z) \label{PropCommutativityAddAsJoinabilityNegationSkolemNormalForm}
\end{eqnarray}
Indeed, $\SStructure$ is a model of $\ol{\cR}\cup\{(\ref{PropCommutativityAddAsJoinabilityNegationSkolemNormalForm})\}$.
But we \emph{should not} conclude (as suggested by the aforementioned sentences in \MaceFour\ manual)
that $\Fadd$ is \emph{not} commutative!
\end{example}
The problem in Example \ref{ExAddMulCommutativity} is that $h:\cI_{\ol{\cR}}\to \SStructure$ is \emph{not surjective}.
For instance, no ground term $t\in\GSTerms$ satisfies $t^\SStructure=1$; note that $\verb!c1!,\verb!c2!\notin\SSymbols$.
Since proving validity in $\cI_\cS$ is not the main purpose of $\MaceFour$, 
no warning in its 
documentation prevents the prospective user to give credit to the `refutation' of 
(ground) commutativity for the addition computed by \MaceFour.
We believe that our work is helpful  to clarify the use of such tools, and even improve it by adding (for instance)
sentences reinforcing surjectivity to avoid the problem discussed above.
For instance, \MaceFour\ obtains no model of 
$\ol{\cR}\cup\SurHomTh^T\cup\{(\ref{PropCommutativityAddAsJoinabilityNegationSkolemNormalForm})\}$
with, e.g., $T=\{\Fz,\Fs(\Fz)\}$.

\subsubsection*{Proofs by Satisfiability vs.\ Theorem Proving.}

In order to further clarify the differences between our
approach and the use
of  first-order theorem proving tools, consider the CTRS $\cR$ in 
\cite[Example 1]{Lucas_AnalysisOfRewritingBasedSystemsAsFirstOrderTheories_LNCS_LOPSTR17}, consisting
of the rules\\[-0.6cm] 
\begin{tabular}{c@{\hspace{-0.6cm}}c}
\begin{minipage}[t]{.51\linewidth}
\begin{eqnarray}
\Fb & \to & \Fa\label{Ex1_LNCS_LOPSTR17_rule1}
\end{eqnarray}\nonumber
\end{minipage} & 
\begin{minipage}[t]{.537\linewidth}
\begin{eqnarray}
\Fa & \to & \Fb \IF \Fc \to \Fb\label{Ex1_LNCS_LOPSTR17_rule2}
\end{eqnarray}\nonumber
\end{minipage}
\end{tabular}\\[0.3cm]
Its associated Horn theory $\ol{\cR}$ 
is:
\\[-0.6cm] 
\begin{tabular}{c@{\hspace{-0.6cm}}c}
\hspace{-1cm}
\begin{minipage}[t]{.58\linewidth}
\begin{eqnarray}
(\forall x)\: x & \to^* & x\label{Ex1_LNCS_LOPSTR17_HornTh_reflexivity}\\
(\forall x,y,z)\: x \to y \wedge y \to^* z \Rightarrow x & \to^* & z\label{Ex1_LNCS_LOPSTR17_HornTh_transitivity}
\end{eqnarray}\nonumber
\end{minipage} & 
\begin{minipage}[t]{.54\linewidth}
\begin{eqnarray}
\Fb & \to & \Fa\label{Ex1_LNCS_LOPSTR17_HornTh_rule1}\\
\Fc \to^* \Fb \Rightarrow \Fa & \to & \Fb \label{Ex1_LNCS_LOPSTR17_HornTh_rule2}
\end{eqnarray}\nonumber
\end{minipage}
\end{tabular}\\[0.3cm]
We consider some simple tests regarding goals $\Fb\to\Fa$ and $\Fa\to\Fb$ and their negations.
We tried such four goals with the following theorem provers: \AltErgo,\footnote{\url{https://alt-ergo.ocamlpro.com/}} \ProverNine/\MaceFour, \PDLtableau,\footnote{\url{http://www.cs.man.ac.uk/~schmidt/pdl-tableau/}} and \Princess\footnote{\url{http://www.philipp.ruemmer.org/princess.shtml}}
(most of them with a web-interface).
Besides attempting a \emph{proof} of each goal with respect to $\ol{\cR}$, tools
\AltErgo, \MaceFour, and \Princess\ can also generate \emph{models} 
of the negation of the tested goal when the proof attempt fails.
The following table summarizes the results of our test:\\[-0.7cm]
\begin{center}
\begin{tabular}{|c|c||c||*{2}{c|}|*{2}{c|}|*{2}{c|}|*{2}{c|}|}
\hline
& Goal   &  & \multicolumn{2}{c||}{\AltErgo} & \multicolumn{2}{c||}{\MaceFour} & \multicolumn{2}{c||}{\PDLtableau} & \multicolumn{2}{c|}{\Princess}  \\
\hline
\# & $\varphi$ & $\cI_\cR\models\varphi$ & $\ol{\cR}\!\vdash\!\varphi$ & $\SStructure\!\models\!\neg\varphi$ & $\ol{\cR}\!\vdash\!\varphi$ & $\SStructure\!\models\!\neg\varphi$ & $\ol{\cR}\!\vdash\!\varphi$ & $\SStructure\!\models\!\neg\varphi$ & $\ol{\cR}\!\vdash\!\varphi$ & $\SStructure\!\models\!\neg\varphi$\\
\hline
1 & $\Fb\to\Fa$              & \emph{true}   & Y  & N & Y  & N & Y  &  -- & Y  & N\\
2 & $\neg(\Fb\to\Fa)$     & \emph{false} & N & Y & N & Y & N & -- & N & Y\\
\hline
3 & $\Fa\to\Fb$              & \emph{false} &  N & Y  & N & Y  & N & -- & N & Y \\
4 & $\neg(\Fa\to\Fb)$     & \emph{true}  & N & Y & N & Y & N & -- & N & Y\\
\hline
\end{tabular}
\end{center}
Goal $\neg(\Fa\to\Fb)$ in row $4$ is \emph{not} directly proved by any tool.
Indeed, since $\neg(\Fa\to\Fb)$ is \emph{not} a logical consequence of $\ol{\cR}$ (see \cite[Example 2]{Lucas_AnalysisOfRewritingBasedSystemsAsFirstOrderTheories_LNCS_LOPSTR17}), 
$\ol{\cR}\vdash\neg(\Fa\to\Fb)$ does \emph{not} hold.
Our \emph{satisfiability} approach
can be used to formally prove 
 that $\cR$ cannot reduce $\Fa$ into $\Fb$, i.e., that $\cI_\cR\models\neg(\Fa\to\Fb)$ (or $\Fa\not\to_\cR\Fb$) holds: 
from row $3$ we see that $\SStructure\models\neg(\Fa\to\Fb)$  
holds for the models $\SStructure$ of $\ol{\cR}$ computed by 
some of 
the tools.
By Corollary \ref{CoroProofsAsSatisfiabilityPreservation}, the desired conclusion $\Fa\not\to_\cR\Fb$ follows.
Note also that row 4 reports on the ability of some tools to obtain  
models of 
$\Fa\to\Fb$.
However, Corollary \ref{CoroProofsAsSatisfiabilityPreservation} cannot be used to conclude that $\Fa\to_\cR\Fb$ holds (which is
obviously wrong): since $\varphi$ in row $4$
is a \emph{negative} literal, condition (b) in Corollary \ref{CoroProofsAsSatisfiabilityPreservation} must be fulfilled before being
able to conclude $\cI_\cR\models\Fa\to\Fb$ from $\SStructure\models\Fa\to\Fb$ for some model $\SStructure$ of $\ol{\cR}$. 
But this is \emph{not} the case in our test set.

\smallskip
Although Remark \ref{ProofsBySemanticRefutation} explains how an arbitrary program property $\varphi$ can be proved by 
using Corollary \ref{CoroProofsAsSatisfiabilityPreservation} (see also Section \ref{SecDealingWithGeneralSentences}), 
from a practical point of view we better think of our approach as \emph{complementary}
 to the use of first-order proof techniques and tools.
Provability of $\varphi$ (i.e., $\cS\vdash\varphi$) implies that $\cI_\cS\models\varphi$ holds.
Thus, as usual, a proof of $\varphi$ with respect to $\cS$ implies that a program $P$ with $\cS=\ol{P}$ has property $\varphi$.
However, as discussed above, showing that $\cS\vdash\varphi$ or $\cS\vdash\neg\varphi$ holds is often impossible. 
We can try to prove $\cI_\cS\models\neg\varphi$ by using Corollary \ref{CoroProofsAsSatisfiabilityPreservation}, though.
For positive sentences  $\varphi$, this is often affordable.

\section{Conclusions and Future Work}\label{SecConclusions}

We have shown how 
to prove 
properties $\varphi$ of computational systems whose semantics can be given
as a first-order theory $\cS$. 
Our \emph{proofs by satisfiability} proceed 
(see Remark \ref{ProofsBySemanticRefutation})
by just
finding a model $\SStructure$ of $\cS\cup\cZ\cup\{\varphi\}$ where $\cZ$  is an 
\emph{auxiliary} theory representing the requirements (a) and (b) in Corollary \ref{CoroProofsAsSatisfiabilityPreservation} (referred to $\neg\varphi$),
so that $\SStructure\models\cS\cup\cZ\cup\{\varphi\}$ implies $\cI_\cS\models\varphi$.
Surjectivity of the interpretation homomorphisms (requirement (a) in Corollary \ref{CoroProofsAsSatisfiabilityPreservation})
is ensured if $\cZ$ includes the appropriate theory $\SurHomTh$ (see Section \ref{SecSurjectiveHomomorphisms});
and requirement (b), for dealing with negative literals, is fulfilled if $\cZ$ includes $\cN$ in Proposition \ref{PropValidityOfArbitraryFormulasWithComplements}.
Our results 
properly subsume the ones in \cite{Lucas_AnalysisOfRewritingBasedSystemsAsFirstOrderTheories_LNCS_LOPSTR17},
which concern \emph{existentially closed boolean 
combinations of atoms} only.
We have also introduced the notion of \emph{refutation witness} which is useful to 
obtain counterexamples by using the symbols in the first-order language rather than values of the computed model.

From a theoretical point of view, the idea of \emph{proving program properties as satisfiability} 
(see Remark \ref{ProofsBySemanticRefutation})
is appealing as it emphasizes the role of \emph{abstraction} (introduced by semantic structures) in theorem proving
and logic-based program analysis.
However, the requirement of surjectivity of the interpretation homomorphisms and the use of theories $\cN$ 
with \emph{negative} information
about some of the predicates 
introduce additional difficulties in the model generation process.
Investigating methods for the practical implementation of our techniques, and also finding specific areas
of application where our approach can be useful (as done in \cite{LucGut_UseOfLogicalModelsForProvingInfeasibilityInTermRewriting_IPL18}, for instance) is an interesting subject
for future work.

Also, our research suggests that further investigation on the  generation of models for many-sorted theories
that 
combines the use of 
finite and infinite domains is necessary.
For instance, 
 \cite{LucGut_AutomaticSynthesisOfLogicalModelsForOrderSortedFirstOrderTheories_JAR18}
explains how to generate such models by interpreting the sort, function, and predicate symbols  
by using linear algebra techniques. This is implememented in \AGES.
Domains are defined as the solutions of matrix inequalities, possibly restricted to an underlying set of
values (e.g., $\integers$); thus, finite and infinite domains can be obtained as particular cases of the same technique.
Since piecewise definitions are allowed, we could eventually provide fully detailed descriptions of functions and predicates
by just adding more pieces to the interpretations.
However,
such a flexibility is expensive.
In contrast, \MaceFour\ is based on a different principle (similar to \cite{KimZha_ModGenTheoremProvingByModelGeneration_AAAI94})
and it is really fast, but only finite domains can be generated. 
This is a problem, for instance, when using Proposition \ref{PropSurjectiveHomomorphismWithOrderNat} to guarantee surjectivity of
homomorphisms $h_s:\GSTerms_s\to\SStructure_s$. Even though $\SStructure_s$ is finite, we still need to be able to interpret $\SortNat$
as $\naturals$, which is not possible with \MaceFour.
For this reason, the examples in Section \ref{SecExampleOfApplication} (where the computed structures $\SStructure$ 
have finite domains for the `proper' sorts \verb$N$, \verb$LN$, and $\genSort$, and only $\SortNat$ is interpreted as an infinite set) 
could not be handled with \MaceFour, or with similar tools
that are able to deal with sorts (e.g., \SEM\ \cite{ZhaZha_GeneratingModelsWithSEM_IJCAR96} or 
the work in \cite{RegSudVor_FindingFiniteModelsInMultiSortedFirstOrderLogic_SAT16}) but which generate finite domains only.

\medskip
\noindent
{\bf Acknowledgements.} I thank the anonymous referees for their comments and 
suggestions. I also thank Philipp R\"ummer and Mohamed Iguernlala for their clarifying remarks about the use of \Princess\
and \AltErgo, respectively.

\end{document}